\documentclass{aastex}
\usepackage{natbib}
\usepackage{graphicx}
\begin{document}

\title{The Heating of Mid-Infrared Dust in the Nearby Galaxy M33: A Testbed for Tracing Galaxy Evolution}
\author{Marie D. Calapa, Daniela Calzetti}
\affil{Department of Astronomy, University of Massachusetts}
\affil{710 North Pleasant Street, Amherst, MA 01003, USA}
\email{mcalapa@umass.edu, calzetti@astro.umass.edu}
\and
\author{Bruce T. Draine}
\affil{Princeton University Observatory}
\affil{Peyton Hall, Princeton, NJ 08544-1001, USA}
\email{draine@astro.princeton.edu}
\author{M\'{e}d\'{e}ric Boquien}
\affil{Institute of Astronomy, University of Cambridge, Madingley Road, Cambridge,
CB3 0HA, UK}
\email{mboquien@ast.cam.ac.uk}
\author{Carsten Kramer}
\affil{Instituto de Radio Astronom'a Milim\'{e}trica (IRAM), Avenida Divina Pastora 7, Local 20, 18012 Granada, Spain}
\email{kramer@iram.es}
\author{Manolis Xilouris}
\affil{Institute for Astronomy, Astrophysics, Space Applications \& Remote Sensing, National Observatory of Athens, P. Penteli, 15236, Athens, Greece}
\email{xilouris@astro.noa.gr}
\author{Simon Verley}
\affil{Dept. de F\'{i}sica Te\'{o}rica y del Cosmos, Facultad de Ciencias, Universidad de Granada, Spain}
\email{simon@ugr.es}
\author{Jonathan Braine}
\affil{Univ. Bordeaux, Laboratoire dÕAstrophysique de Bordeaux, F-33270, Floirac, France}
\email{braine@obs.u-bordeaux1.fr}
\author{Monica Rela\~{n}o-Pastor}
\affil{Dep. F\'{i}sica Te\'{o}rica y del Cosmos, Campus de Fuentenueva, Universidad de Granada, 18071 Granada, Spain}
\email{mrelano@ugr.es}
\author{Paul van der Werf}
\affil{Leiden Observatory, Leiden University, P.O. Box 9513, NL-2300 RA Leiden, The Netherlands}
\email{pvdwerf@strw.leidenuniv.nl}
\author{Frank Israel}
\affil{Sterrewacht Leiden, Leiden University, PO Box 9513, 2300 RA, Leiden, The Netherlands}
\email{israel@strw.leidenuniv.nl}
\author{Israel Hermelo}
\affil{Departamento de F\'{i}sica Te\'{o}rica y del Cosmos, Universidad de Granada, Spain}
\email{hermelo@iram.es}
\author{Marcus Albrecht}
\affil{Argelander-Institut fŸr Astronomie, Auf dem HŸgel 71, 53121 Bonn, Germany}
\email{malbrecht@astro.uni-bonn.de}

\begin{abstract}
\noindent
Infrared emission is an invaluable tool for quantifying star formation in galaxies. Because the 8 $\mu$m polycyclic aromatic hydrocarbon (PAH) emission has been found to correlate with other well-known star formation tracers, it has widely been used as a star formation rate (SFR) tracer. There are, however, studies that challenge the accuracy and reliability of the 8 $\mu$m emission as a SFR tracer.\\
\indent
Our study, part of the Herschel\footnote{Herschel is an ESA space observatory with science instruments provided by European-led Principal Investigator consortia and with important participation from NASA.} M33 Extended Survey (HERM33ES) open time key program, aims at addressing this issue by analyzing the infrared emission from the nearby spiral galaxy M33 at the high spatial scale of $\sim$ 75 pc. Combining data from the \it{Herschel} \rm Space Observatory and the \it{Spitzer} \rm Space Telescope we find that the 8 $\mu$m emission is better correlated with the 250 $\mu$m emission, which traces cold interstellar gas, than with the 24 $\mu$m emission. Furthermore, the L(8)/L(250) ratio is more tightly correlated with the 3.6 $\mu$m emission, a tracer of evolved stellar populations and stellar mass, than with a combination of H$\alpha$ and 24 $\mu$m emission, a tracer of SFR. The L(8)/L(24) ratio is highly depressed in 24 $\mu$m luminous regions, which correlate with known HII regions.\\ 
\indent
We also compare our results with the dust emission models by \citet{draine07}. We confirm that the depression of 8 $\mu$m PAH emission near star-forming regions is higher than what is predicted by models; this is possibly an effect of increased stellar radiation from young stars destroying the dust grains responsible for the 8 $\mu$m emission as already suggested by other authors. We find that the majority of the 8 $\mu$m emission is fully consistent with heating by the diffuse interstellar medium, similar to what recently determined for the dust emission in M31 by \citet{draine13}. We also find that the fraction of 8 $\mu$m emission associated with the diffuse interstellar radiation field ranges between $\sim$60\% and 80\% and is 40\% larger than the diffuse fraction at 24 $\mu$m.
\end{abstract}
\keywords{ISM: dust, extinction; galaxies: individual (M33); galaxies: Local Group; galaxies: ISM; infrared: ISM}

\section{Introduction}

\noindent
In order to understand the evolution of galaxies in the Universe, accurate measures of star formation (SF) within galactic structures need to be obtained. A large number of  tracers of star formation in a galaxy have been defined in the literature, using different regions of the electromagnetic spectrum. Among these, the infrared (IR) radiation is a classical tracer of activity in galaxies \citep{kennicutt98, ken12, calz12}. In star-forming galaxies, young, massive stars are responsible for most of the ultraviolet (UV) radiation and, in the presence of dust a smaller or larger fraction of their light may be absorbed and reradiated in the IR regime. By observing galaxies in IR light, determinations about the rate and areas of star formation can be surmised.\\
\indent
The mid-IR wavelength region ($\approx$7-40 $\mu$m) in general, and the emission around $\sim$8 $\mu$m in particular, are among the favored tracers of recent star formation, because of their ready detectability in galaxies at high redshift (e.g., Daddi et al. 2005, Reddy et al. 2010, 2012). The emission in the  $\sim$8 $\mu$m region is mainly contributed by a combination of stellar photospheric emission, the featureless continuum of hot dust emission, and the Polycyclic Aromatic Hydrocarbon 
(PAH) spectral features (e.g., \citet{smith07}). As the PAH features contribute about 70\% or more of the emission in the Spitzer $\sim$8 $\mu$m band \citep{smith07}, we call the emission in this band `PAH emission' henceforth. \citet{roussel01} and \citet{forster04} have suggested, using data from
 ISO,  that the PAH emission is closely related to the hot dust emission at 15 $\mu$m and other tracers of star formation. Other studies (e.g. \citet{bos04,haas02}) and follow up investigations, utilizing the \it{Spitzer} \rm Space Telescope \citep{werner04}, found the relation to be more complex than previously inferred. The PAH emission does not immediately correlate with star formation as traced by the 24 $\mu$m emission,  as the L(8)/L(24) luminosity ratio is depressed in regions of known star formation relative to the ratio in the diffuse starlight field \citep{helou04, calz05, bendo08, pov07}. \citet{bendo06, bendo08}, furthermore, showed that the PAH emission is more closely related to tracers of cool dust emission, heated by the diffuse starlight field, than tracers of warm dust, heated by recent star formation. This is confirmed by \citet{verley09}, who used a complementary approach to our own to determine that at least 60\% of the 8 and 24 $\mu$m emission in M33 is diffuse. Conversely, a recent study by \citet{crock13} determined that the 8 $\mu$m emission in the galaxy NGC 628 has only a 30\% - 43\% fraction unassociated with recent star formation. These contradictory results call for further studies to assess the validity of the PAH emission as a SFR tracer.\\
\indent
The powerful \it{Herschel} \rm Space Observatory \citep{pilbratt10} has targeted nearby galaxies with unprecedented resolution at the infrared wavelengths ($\geq 70 \mu$m) where the warm dust emission, powered primarily by massive stars, is progressively supplanted by the cold dust emission powered by low mass stars \citep{bendo10,bendo12,boq11}. The synergy between the \it{Spitzer} \rm and \it{Herschel} \rm imaging data is such that the infrared spectral energy distribution of individual star forming regions from $\sim$3 $\mu$m to the sub-mm can be separated from that of the diffuse starlight in nearby galaxies, permitting perusal of the emission originating in these regions. With this level of detail, the properties of dust emission and its relation to the SFR can be better investigated. This paper uses data from both \it{Spitzer} \rm and \it{Herschel} \rm to observe the nearby galaxy, M33, in a range of IR wavelengths.\\
\indent
M33 is a member of the Local Group and a spiral galaxy, with an inclination of 56$^\circ$ \citep{regan94} and a distance of 840 kpc \citep{freedman91}. Because of its proximity, M33 is an ideal site for the investigation of the properties of dust emission: 1$^{\prime\prime}$ subtends a spatial scale of $\sim$4 pc. M33 harbors a large number of HII regions \citep{bou74,hodge99,verley10, rel13}, which are easily identifiable in the \it{Spitzer} \rm  and \it{Herschel} \rm images. M33 has an oxygen abundance of about half solar and a shallow--to--negligible metallicity gradient as a function of galactocentric distance \citep{ros08, magrini09,bre11}; this characteristic enables us to investigate the PAH emission with less attention to the effect of metallicity on the strength of the PAH features \citep{hunt05, eng05, mad06,draine072, eng08}. The \it{Herschel} \rm data for this project came from the HERM33ES open time key program \citep{kramer10} and \it{Spitzer} \rm data were obtained through the Guaranteed Time observations for the IRAC and MIPS instruments \citep{hinz04, mcquinn07}. We also make use of the H$\alpha$ image from \citet{hoopes00}, taken with the 0.6 meter Burrell-Schmidt telescope at Kitt Peak National Observatory. Details about the imaging and its reduction can be found in that paper.\\
\indent
Relationships among the PAH 8 $\mu$m, 24 $\mu$m, 250 $\mu$m, and total infrared (TIR) emission are the primary subjects of this study, under the reasonable assumption that the 24 $\mu$m emission is mainly tracing current SF \citep{calz05,calz07} and that the 250 $\mu$m emission is mainly tracing cold dust.\\
\indent
Our analysis plan is similar to that used by \citet{bendo08} for the SINGS galaxies \citep{ken03}: we will investigate the relationship of the  8 $\mu$m emission with tracers of the warm  (24 $\mu$m) and cool (250 $\mu$m) dust emission, and with the total infrared (TIR) emission. The advantage of our analysis over that of Bendo et al. is twofold: (1) M33 is 3 to 20 times closer than the SINGS galaxies, thus enabling exquisite spatial resolution; and (2) the 250 $\mu$m emission from {\it Herschel} traces more closely the cool dust and has more than twice the spatial resolution of the 160 $\mu$m {\it Spitzer} data used by Bendo et al.\\
\indent
In what follows, L($\lambda$) refers to monochromatic luminosity derived as 
\begin{equation}
L(\lambda) = [\nu L_{\nu}]_{\lambda}
\end{equation}
in units of erg s$^{-1}$ kpc$^{-2}$. In this work, each pixel will be $\sim$73.3 pc in size, or about the size of a large HII region.

\section{Data}

\begin{figure}[ht]
\includegraphics[angle=270,scale=0.7]{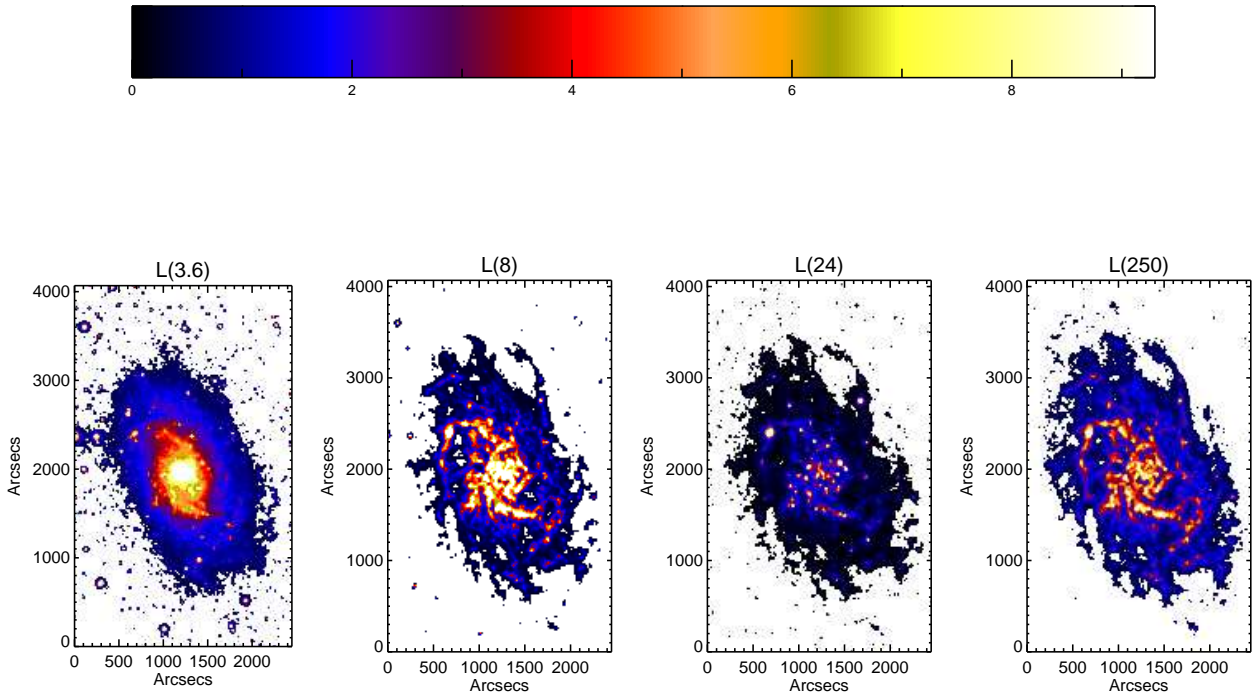}
\caption{Luminosity maps at 3.6, 8, 24, and 250 $\mu$m. Each map has been convolved to the 250 $\mu$m resolution (PSF = 19.7$^{\prime\prime}$) and resampled in bins of 18$^{\prime\prime}$ $\times$ 18$^{\prime\prime}$. The units of the color bar are 10$^{40}$ erg s$^{-1}$ kpc$^{-2}$.}
\end{figure}

\noindent
This study makes use of mosaics of M33 obtained in the following bands: 3.6, 8, 24, 70, 160, and 250 $\mu$m. Figure 1 shows luminosity maps of 3.6, 8, 24, and 250 $\mu$m images. The 3.6, 8, 24, and 70 $\mu$m images were obtained with the Infrared
Array Camera \citep[IRAC for 3.6 and 8 $\mu$m, ][]{fazio04} and the
Multiband Imaging Photometer for Spitzer \citep[MIPS for 24 and 70 $\mu$m, ][]{rieke04} instruments on the \it{Spitzer} \rm Space Telescope \citep{werner04}. The PSF of the images ranges from $\sim$2$^{\prime\prime}$ to $\sim$18$^{\prime\prime}$. The 3.6
$\mu$m image mainly contains photospheric emission from stars \citep{meidt12}, and
was used to subtract stellar contributions in the 8 and 24 $\mu$m images.\\
\indent
Emission at 24 $\mu$m in nearby galaxies comes primarily from hot dust ($\gtrsim$ 100 K), and becomes luminous more rapidly than other IR bands in strong radiation fields, such as star forming regions \citep{draine07}. It has been calibrated as a SFR tracer, either by itself or in combination with H$\alpha$ \citep{herrero05, calz05,calz07}. 70 $\mu$m is also a reasonable tracer of star forming regions in luminous galaxies, as it accounts for a significant fraction of the total IR emission ($\sim$30\% - 50\% ; \citet{calz10, boq10, li10,li13}). The 160 $\mu$m image was obtained using the Photodetector Array Camera and Spectrometer (PACS; \citet{pog10}) on the \it{Herschel Space Observatory} \rm \citep{pilbratt10}. These bands were used to derive the TIR image of M33, as they trace the bulk of the total infrared emission. The 250 $\mu$m image was obtained using the Spectral and Photometric Imaging Receiver (SPIRE; \citet{griffin10}) on \it{Herschel} \rm. For more details about the reduction and nature of the data, see \citet{kramer10, verley10,boq11}.\\
\indent
The images were convolved to the PSF of the 250 $\mu$m image, 19.7$^{\prime\prime}\pm$ 1.7$^{\prime\prime}$ (Kramer et al. 2010). This was done using a convolution script in IDL and kernels from \citet{aniano11}. Using Fourier transformations, this process takes an image, resamples it, and resizes it to match a reference kernel. When this was completed, the image was aligned to the 250 $\mu$m image. Thus, all images used for our analysis were processed to be at the same pixel size, aligned with, and with the same field of view as the 250 $\mu$m image.\\
\indent
Next, we subtracted the stellar component of the infrared emission in the 8 and 24 $\mu$m
bands. In order to find the optimal factor of the 3.6 $\mu$m emission to subtract, the fluxes of 50 point 
sources (foreground stars) were measured in all three images: 3.6, 8, and 24 $\mu$m. Histograms of the 8/3.6 and 24/3.6 flux ratios were used to identify the peak ratio value, to be used as initial guess for the optimal fraction of the 3.6 $\mu$m fluxes to be subtracted from each of the 8 and 24 $\mu$m images, respectively. With these initial guesses, we produced several images with varying degrees of stellar subtraction, which were compared to establish the fractions of 3.6~$\mu$m fluxes that would yield the least amount of stellar residual, while at the same time avoiding over-subtraction in the regions of diffuse emission. We found these fractions to be 0.35 for the 8~$\mu$m image and 0.06 for the 24~$\mu$m image, with all images still in units of MJy/sr. Although these fractions are larger than 
those derived by Helou et al (2004) and Calzetti et al. (2007), they are consistent with expectations 
from stellar population synthesis models with star formation durations $\approx$1~Gyr or longer 
(Leitherer et al. 1999). Choosing Helou et al.'s (2004) factor of 0.25 for the rescaling of the 3.6 $\mu$m image, instead of our derived 0.35, changes the measured 8 $\mu$m fluxes by 8\% on average, as derived from the mode of the ratio of the two 8 $\mu$m images. Furthermore, we performed checks on the stellar-continuum subtracted 8 and 
24 $\mu$m images at the original (non-convolved) resolution, to verify that the subtraction of the 
3.6~$\mu$m image is effective at removing also the contribution of foreground stars that are located
within the extent of the galaxy, while ensuring that no region is over-subtracted.\\
\indent
In order to improve S/N, the images were rebinned to be pixels 18$^{\prime\prime}$ on a side. At the distance of M33, this angular resolution corresponds to a spatial scale of 73 pc, which is the size of a large HII region. Thus, our re-binning of the images does not impact our results, since we can
still easily separate regions dominated by young stars from more
quiescent regions. Rebinning to the size of the 250 $\mu$m PSF also makes our pixels independent.\\
\indent
After stellar subtraction and rebinning, sky background was removed from all images, using the msky procedure available in IRAF\footnote{Dr. Mark Dickinson, NOAO, private communication}, which calculates the mode and the standard deviation of the average pixel values. We use the standard deviation as our 1-$\sigma$ value for that band. Only fluxes above the 5-$\sigma$ level are used in the sky subtracted images, to ensure that measurements are significant; with this threshold selection, we usually have at least 4500 separate pixels in each of the images.\\
\indent
Extended source aperture corrections were applied to the 3.6 and 8 $\mu$m data due to the extended nature of the features in M33. To account for this, the 3.6 $\mu$m flux was multiplied by a constant of 0.96 and the 8 $\mu$m flux was multiplied by a constant of 0.81, consistent with the correction to an infinite aperture photometry of our pixel photometry, as found in the IRAC Instrument Handbook\footnote{irsa.ipac.caltech.edu/data/SPITZER/docs/irac/iracinstrumenthandbook/}.\\
\indent
A total infrared (TIR) surface brightness image was produced for comparison with the monochromatic images and ratio images. Following the prescription of \citet{dale02}, the single-band images of M33 are added to produce a TIR image in the following way:
\begin{equation}
\rm L_{\rm (TIR)}\rm = 1.559 L(24) + 0.7686 L(70) + 1.347 L(160)
\end{equation}
where L($\lambda$) is defined as in Equation 1.
The luminosities determined in Equation 1 are found to be in agreement with those determined using a more extensive range in fluxes, as derived in \citet{boq11} paper, typically within 4 - 5\% in all bins with a dispersion of 5 - 10\%. The recipe given in \citet{draine07}, equation 22, is similar to equation 2 above, but includes the 8 $\mu$m emission in the TIR estimate. Comparisons between maps using our equation 2 and the \citet{draine07} estimator show a systematic offset of 8\% (equation 2 giving a lower TIR estimate) with a dispersion of 7\%. Given the similarity of all estimators and the absence of systematic differential trends across the galaxy, we adopt equation 2 for our analysis. Figure 2 shows the TIR luminosity map of the galaxy. In this Figure, the TIR maps shows a larger fraction of bins below our S/N=5 cut than any of the images shown in Figure 1. This is due to the use of the lower S/N
map at 160 $\mu$m from \textit{Herschel}/PACS to produce the TIR image.
\begin{figure}[ht]
\centering
\includegraphics[angle=270,scale=0.7]{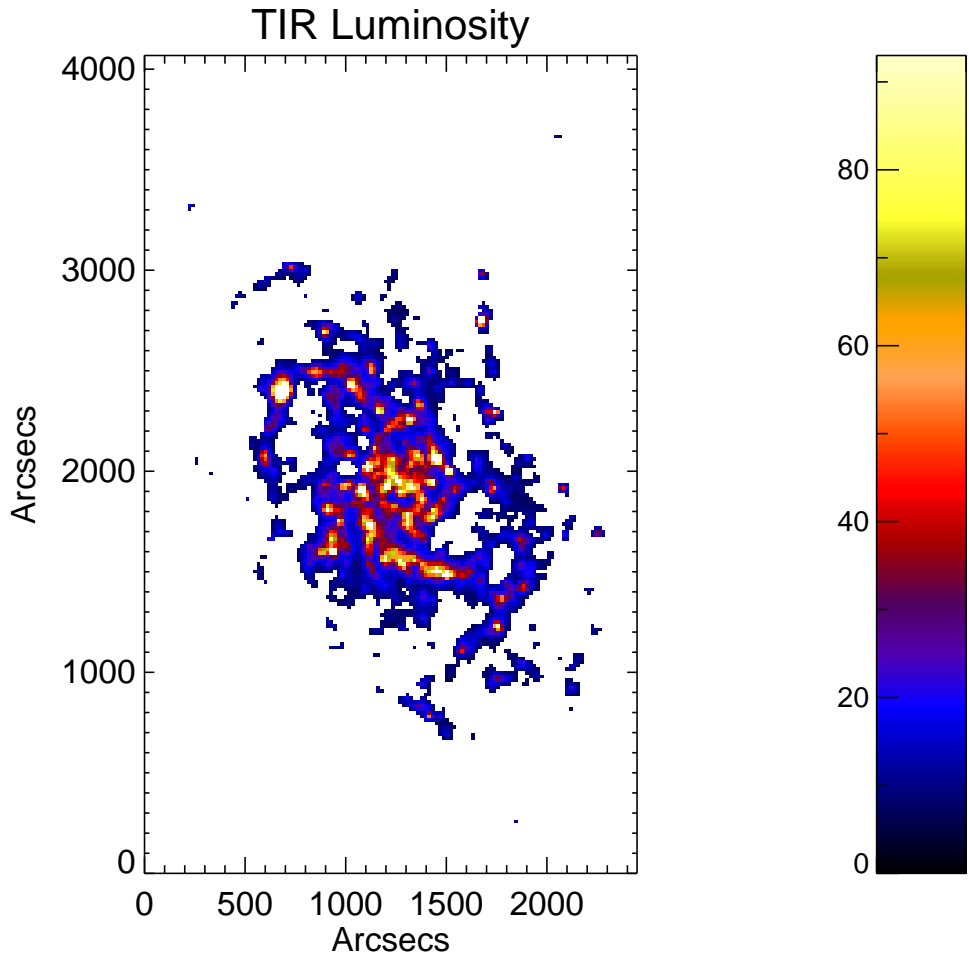}
\caption{TIR Luminosity map image. The resolution of the image is the same as that reported in Figure 1. The units of the color bar are 10$^{40}$ erg s$^{-1}$ kpc$^{-2}$.}
\end{figure}

\section{Analysis}

%\subsection{On 5-$\sigma$ Boundaries in Plots}

%\begin{figure}
%\epsscale{0.7}
%\plotone{bounds.eps}
%\caption{The scatter plot of the L(8)/L(24) ratio versus the 250 $\mu$m luminosity plot, for the pixels (18$^{\prime\prime}$ $\times$ 18$^{\prime\prime}$ = 73.3 $\times$ 73.3 pc $^{2}$) above 5-$\sigma$ in the M33 images. Each data point in thefigure corresponds to one pixel. Blue contours on the data show the density of points in an area of the plot. Representative 1-$\sigma$ error bars are shown in red along the bottom of the data. A running median is plotted as a yellow line and points with orange dispersion bars. The 5-$\sigma$ limits are shown as the blue lines. The upper and lower horizontal boundaries were calculated by taking the 5-$\sigma$ limit on one image and the maximum flux value measured in the other image. These limits are much larger than the typical luminosity ratios measured in this work and do not impose a limitation in our analysis.}
%\end{figure}
%\noindent

\subsection{Models}

\begin{figure}
\epsscale{0.7}
\plotone{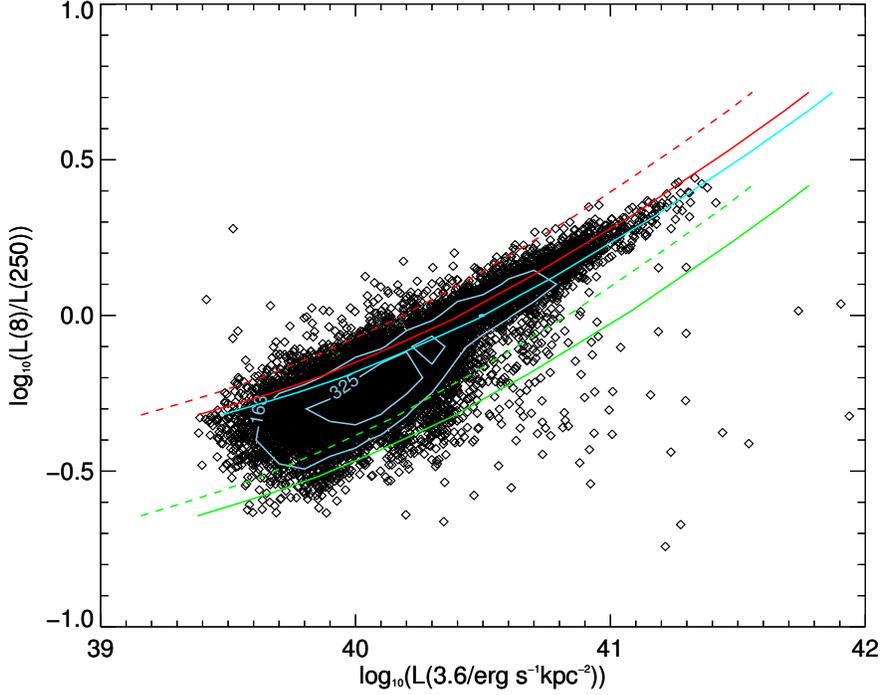}
\caption{Scatter plot showing L(8)/L(250) luminosity ratio versus L(3.6), the stellar luminosity in the 3.6 $\mu$m IRAC band for the pixels in the M33 images. The red line is the \citet{draine07} q$_{PAH}$ = 0.046 model with an $f$ factor of 1.25 (solid line, see equation 3) and 0.75 (dashed line), respectively. The green line is the 2.5\% PAH mass fraction case with an $f$ factor of 1.25 (solid line, see equation 3) and 0.75 (dashed line), respectively. The $f$ value and PAH mass fraction adopted in this paper is shown as the solid cyan line, q$_{PAH}$ = 0.046 with $f$ = 1.55. Models are plotted for U$_{min}$ = U$_{max}$ (singleÑvalued ISRF), where
U$_{min}$ covers the range 0.1-25. Contours in all figures report the number of points within a contour region. The model L(3.6) is calculated as described in Section 3.2.}
\end{figure}
\noindent
We compare our data with the dust emissivity models of \citet{draine07}. Briefly, these are physically-motivated models that describe the heating of dust by starlight, where 
the dust is a mixture of amorphous silicate and graphitic grains, including PAH particles. The fraction of PAHs present in the models is a parameter called 
q$_{PAH}$, which is the percentage by mass of PAHs relative to the total dust mass. A number of authors \citep{hunt05,eng05,mad06,draine072,smith07,eng08} have established that above a critical metallicity value of about 20\% solar, the value of 
q$_{PAH}$ is relatively constant, and close to the Milky Way value of 0.04. Here, we adopt the closest value for which models are available, q$_{PAH}$=0.046, 
as our default for M33, on account of this galaxy being only about 0.5 solar in metal abundance.In some cases, for comparison, we also present the q$_{PAH}$=0.025 models. This is noted in the figure captions.\\
\indent
In the \citet{draine07} models, the starlight intensity heating the dust is modeled as a dimensionless factor $U$ that multiplies the general interstellar 
radiation field (ISRF) in the solar neighborhood, as estimated by \citet{mathis83}. In order to account for the range of starlight intensities present in galaxies, 
these authors parametrize the starlight as the sum of two contributions: one describing the ISRF of the galaxy, with dimensionless factor $U_{min}$, and the 
other describing regions of higher starlight intensity, with $U$ varying between U$_{min}$ and U$_{max}$. The latter factor accounts for the heating of  dust by 
the increasingly higher stellar energy densities in the proximity of OB associations and other high energy regions, where $U>100$. More details on the models, and the predicted dust emission intensity in the various Spitzer and Herschel bands can be found in the original paper  \citep{draine07}. Here we will compare our 
spatially--resolved data of M33 with these models, for varying U$_{min}$ and U$_{max}$.\\ 
\indent 
We perform comparisons using both luminosity ratios (e.g., L(8)/L(24)) and luminosities (e.g., L(3.6), L(24), etc.). Luminosity ratios are readily available 
from the models, while the model luminosities need to be derived for our specific case (distance, region size, etc.). The model stellar luminosity at 3.6 $\mu$m 
is obtained assuming that the ISRF spectrum of \citet{mathis83} for the solar neighborhood is applicable to M33. We rescale this spectrum to the distance of 
M33 and our region sizes of 73.3pc. In order to account for the fact that we are observing M33 from an external vantage point, rather than from within the 
galaxy itself (as if the case for the Milk Way), we model the galaxy as a plane parallel homogeneous distribution of stars and dust, with inclination of 
56$^{o}$, yielding the following `net' expression for the 3.6~$\mu$m luminosity per pixel:
\begin{equation}
\rm{L(3.6) = <U> \times f \times 1.922 \times 10^{40}\ \  erg s^{-1}kpc^{-2}}
\end{equation}
where $<U>$ is the mean value of the starlight intensity in the pixel, and $f$ is a factor that accounts for variations in the
dust optical depth (thus, gas column densities) in the disk. For a metal abundance about 0.5 solar and a total gas, HI$+$H$_2$, gas column density in the range 
4$\times$10$^{20}$ -- 6$\times$10$^{21}$~cm$^{-2}$ \citep{gratier10, braine10}, $f$ is in the range [0.75; 1.25]. See the Appendix for the derivation of equation 3. Figure 3 shows the scatter plot 
of L(8)/L(250) versus L(3.6) for the M33 pixels\footnote{In producing the plots that follow, it was found that the 5-$\sigma$ boundaries on the emission ratios usually lay far from the data points themselves. In this case, the 5-$\sigma$ lines do not provide a selection bias. For this reason, we have chosen to plot the 5-$\sigma$ boundaries only in the cases for which they provide a selection limit on the data.}, with the expectations from the \citet{draine07} model overlaid, for $f$=0.75 and $f$=1.25; for  the model, we adopt both q$_{PAH}$=0.046 (see above) and the lower value q$_{PAH}$=0.025 for completeness. We allow $<U>=U_{min}$ to span the 
range 0.1 to 25. This range is in agreement with what \citet{draine13} find for the Andromeda galaxy. We find a small offset in Figure 3 between the 
q$_{PAH}$=0.046 model and the data, which can be caused by any number of the following reasons: (1) the ISRF spectrum of M33 may differ from that of the 
Milky Way; (2) the dust and stars are not homogeneously distributed across the disk; (3) dust and stars may have different scale heights in the galaxy, and these 
scales may also depend of the mean stellar population age; (4) the fraction of PAHs, q$_{PAH}$, in M33 may be slightly different (lower) from that of the Milky Way, although our data do not have the accuracy to enable an exact determination of this parameter. As in the case of L(3.6), the dust luminosities depend on both the
starlight intensity $<$U$>$ (Draine \& Li 2007) and the range of gas
column densities (dust opacities) parametrized by the factor $f$,
which we adopt to have fixed value $f\sim 1.55$. Thus, for the purpose of this analysis, we adopt q$_{PAH}$=0.046 and a mean value of $f$ that approximately matches the data in Figure 3, $f\sim$1.55, and keep it constant in the rest of the paper. Where relevant, we also show the case q$_{PAH}$=0.025.\\
\indent
The model dust luminosities, L(8), L(24), L(250), etc., are derived by converting the emissivity per
unit H atom to a total emissivity per pixel by using the range of  HI$+$H$_2$ column densities given above, after multiplying said emissivity by 0.5 to account 
for the roughly half solar metallicity of M33, and recalling that each pixel in our analysis 
subtends 73.3 pc. As in the case of L(3.6), the dust luminosities are also parametrized by $<U>$, and are multiplied by the same factor $f\sim$1.55. In order to account for the range of gas column densities
observed in M33, we show in all relevant figures the model
dust luminosities derived under the two
extreme assumptions of N(HI+H$_2$)=6$\times$10$^{21}$~cm$^{-2}$
and N(HI+H$_2$)=4$\times$10$^{20}$~cm$^{-2}$.\\

\subsection{Relationships Between 8 $\mu$m and 24 $\mu$m Emission}

\begin{figure}
\centering
\includegraphics[scale=0.7]{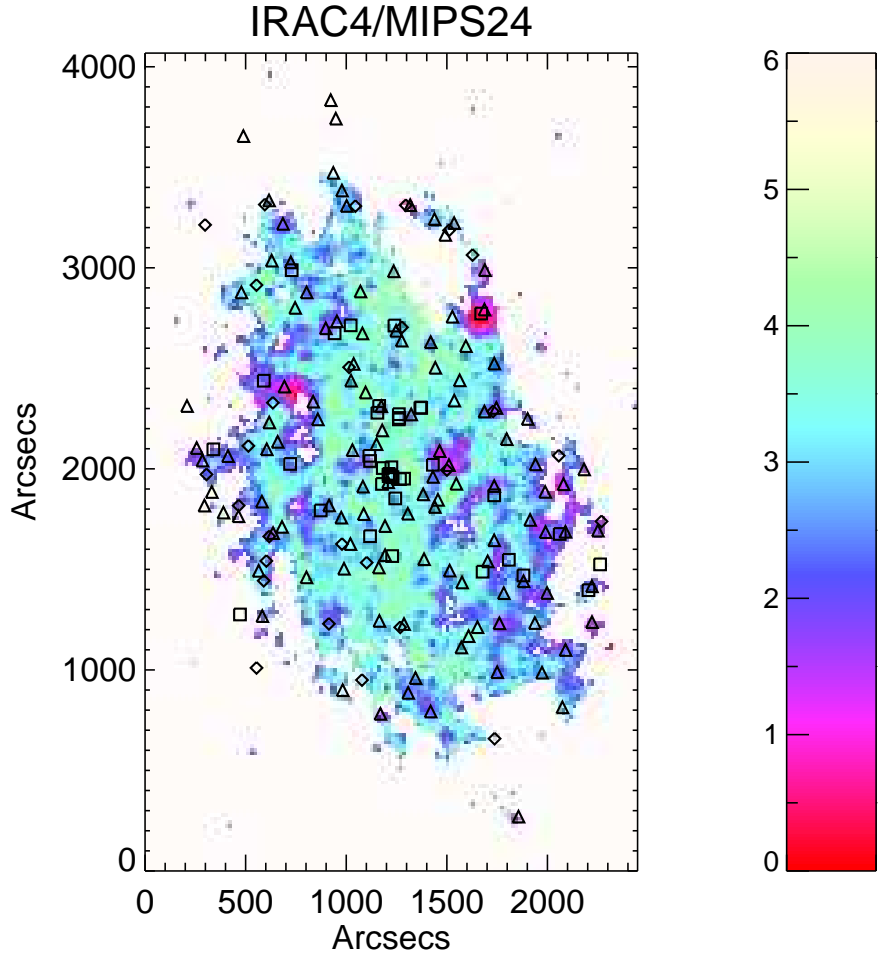}
\caption{The ratio image of L(8)/L(24) luminosity. The resolution of the image is the same as that reported in Figure 1. Prominent HII regions are identified in the image; 27 regions from \citet{verley07} are denoted by diamonds, 119 regions from \cite{rel13} are denoted by triangles, and 39 of the brightest HII regions from \citet{hodge99} are denoted by squares. HII regions tend to cluster close to the minima in the L(8)/L(24) ratio.}
\end{figure}
\noindent
Figure 4 shows the ratio map of L(8)/L(24) luminosity of the galaxy. The ratio is somewhat homogenous over the galaxy, with minima in documented HII regions \citep{hodge99, verley07,rel13}. One prominent example is seen in the upper right arm, where a large minimum is present. This minimum correlates with a luminous region in the 70 $\mu$m image, which traces star forming regions. The presence of L(8)/L(24) ratio minima in HII regions indicates that 8 $\mu$m emission is under luminous within HII regions, as already noted by \citet{helou04} in the galaxy NGC 300 and \citet{pov07} in the Milky Way HII region M17. This trend is better seen in Figure 5, where the L(8)/L(24) luminosity ratio is plotted as a function of the 24 $\mu$m luminosity (in each pixel), showing that luminous regions correspond to low L(8)/L(24) ratio values.\\

\begin{figure}
\epsscale{0.8}
\plotone{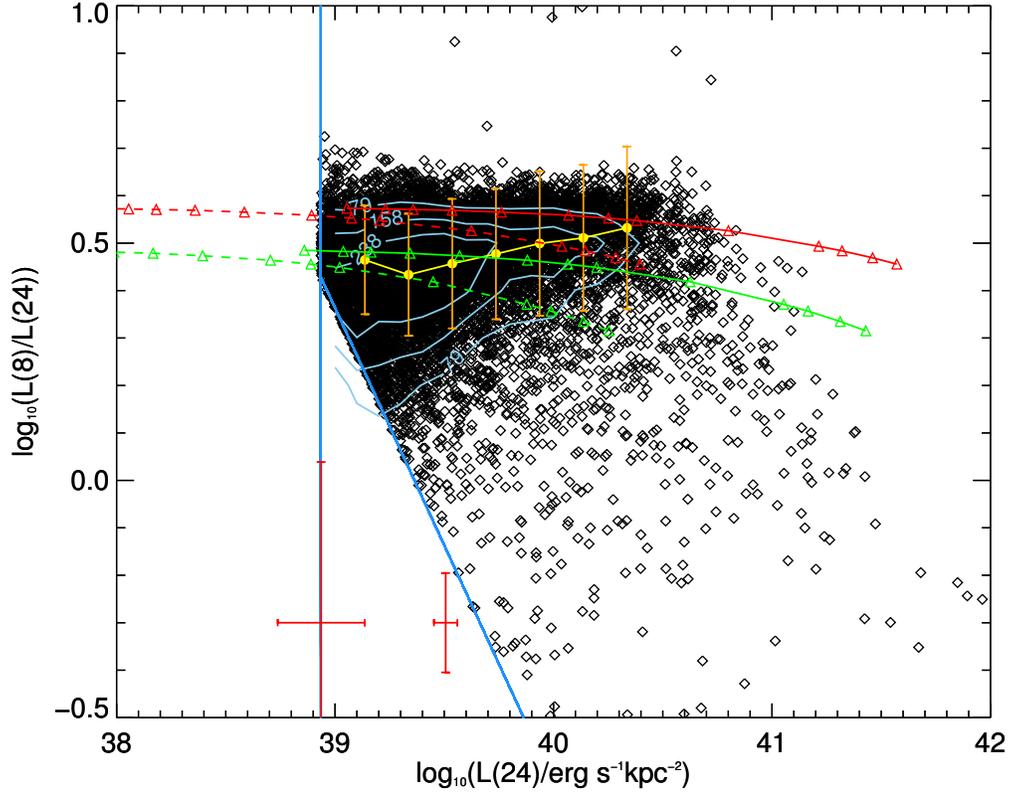}
\caption{The ratio of L(8)/L(24) versus 24 $\mu$m luminosity plot. Blue lines mark the 5-$\sigma$ limit on the data. The median values of the data are presented as the yellow points and line, with dispersion bars at each point the median was calculated. The predicted values for the emissivity from \citet{draine07} for the q$_{PAH}$ = 0.046 case (red lines; section 3.2) and the q$_{PAH}$ = 0.025 case (green lines), are plotted for U$_{min}$ = U$_{max}$ in the range 0.1 - 25. Dashed lines are models calculated using an N$_{H}$ of 4 $\times$ 10$^{20}$ cm$^{-2}$ and solid lines are models calculated using an N$_{H}$ of 6 $\times$ 10$^{21}$ cm$^{-2}$. Representative red 1-$\sigma$ uncertainty bars on the data are presented at the bottom of the plot.}
\end{figure}
\indent
In Figure 5, a generally flat trend of the data is evident, confirmed by the running median of the data points. This is consistent with the range of slopes found in other galaxies analyzed by \citet{bendo08}. In their paper, such a relationship as is found in M33 was taken to imply a dissociation between the 8 and 24 $\mu$m emission. One difference between our findings and those of Bendo et al. should be noted, however. In their paper, they report a systematically decreasing trend of the 8/24 ratio for all 24 $\mu$m surface brightnesses, even at the faintest level. This disparity between our results and those of Bendo likely stems from our much improved spatial resolution (73 parsecs versus the typical 550 parsecs of Bendo et al., i.e., a factor 7.5 better spatial resolution), that enables us to cleanly separate regions of current star formation from the more quiescent regions.\\
\indent
Our data are flatter overall until a log$_{10}$(L(24)/erg s$^{-1}$ kpc$^{-2}$) value of $\sim$40.65, after which a sharp decline in the L(8)/L(24) ratio occurs. For our resolution, this corresponds to a total 24 $\mu$m luminosity of  2.1 $\times$ 10$^{38}$ erg s$^{-1}$, which would correspond to an H$\alpha$ luminosity of 6.5 $\times$ 10$^{36}$ ergs$^{-1}$, when using equation 5 of Calzetti et al. 2007. This is the H $\alpha$ luminosity of an HII region powered by a 5 Myr old star cluster with mass of about 800 M$_{\odot}$ (Starburst99 models; \citet{lei99}). These are some of the smallest star clusters that can dominate the emission within regions of 70 pc size.\\
\indent
The models over-plotted on the data are for both cases of q$_{PAH}$ = 0.046 and q$_{PAH}$ = 0.025 with U$_{min}$ = U$_{max}$ and ranges from 0.1 to 25. Two different line styles are shown in Figure 5, corresponding to two different assumptions for the column density of total hydrogen in M33. Values for N$_{H}$ are from the column density maps of \citet{gratier10} and \citet{braine10}, which describe an approximate range of N$_{H}$ from 4 $\times$ 10$^{20}$ - 6 $\times$ 10$^{21}$.\\

\begin{figure}[h]
\epsscale{1}
\plottwo{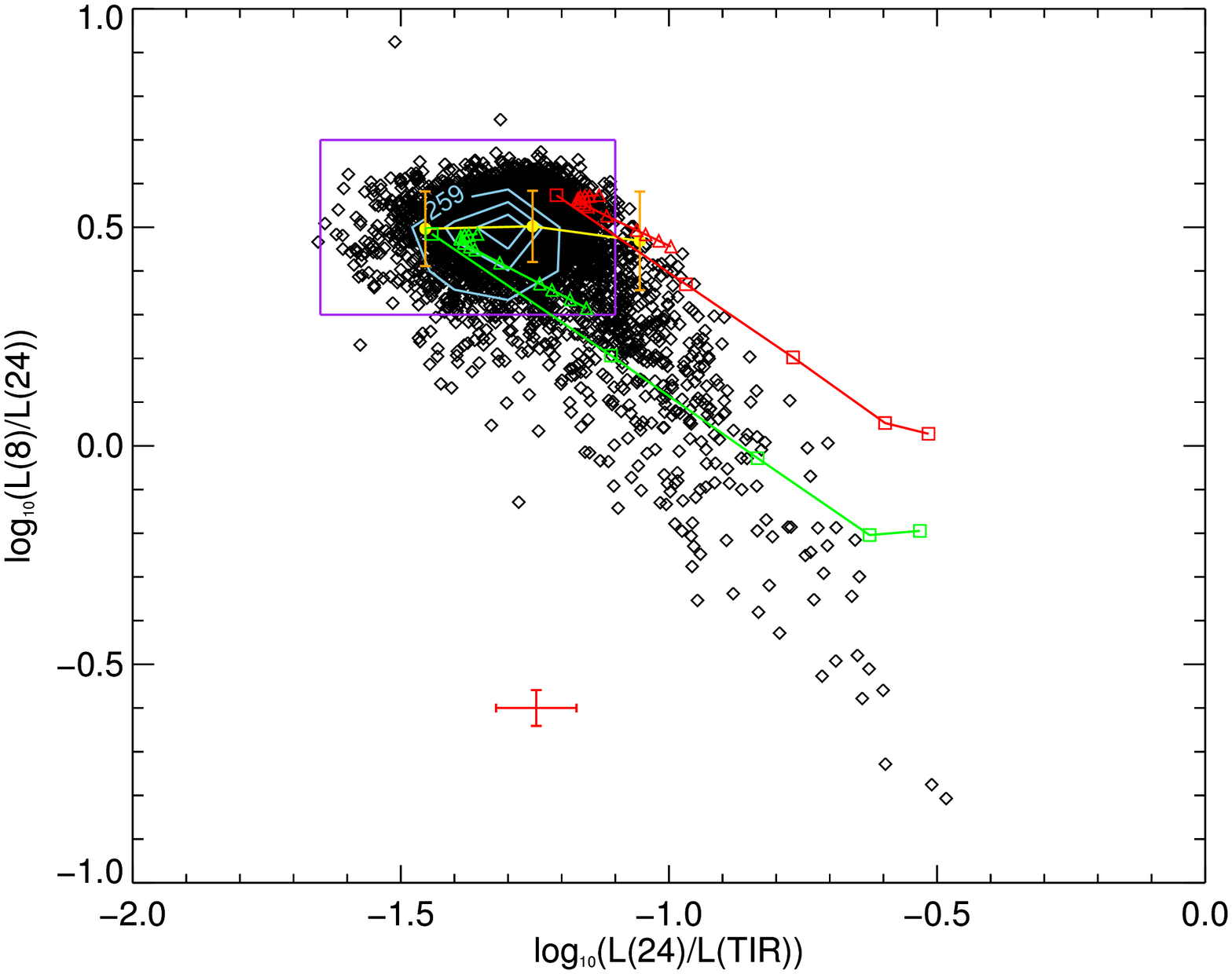}{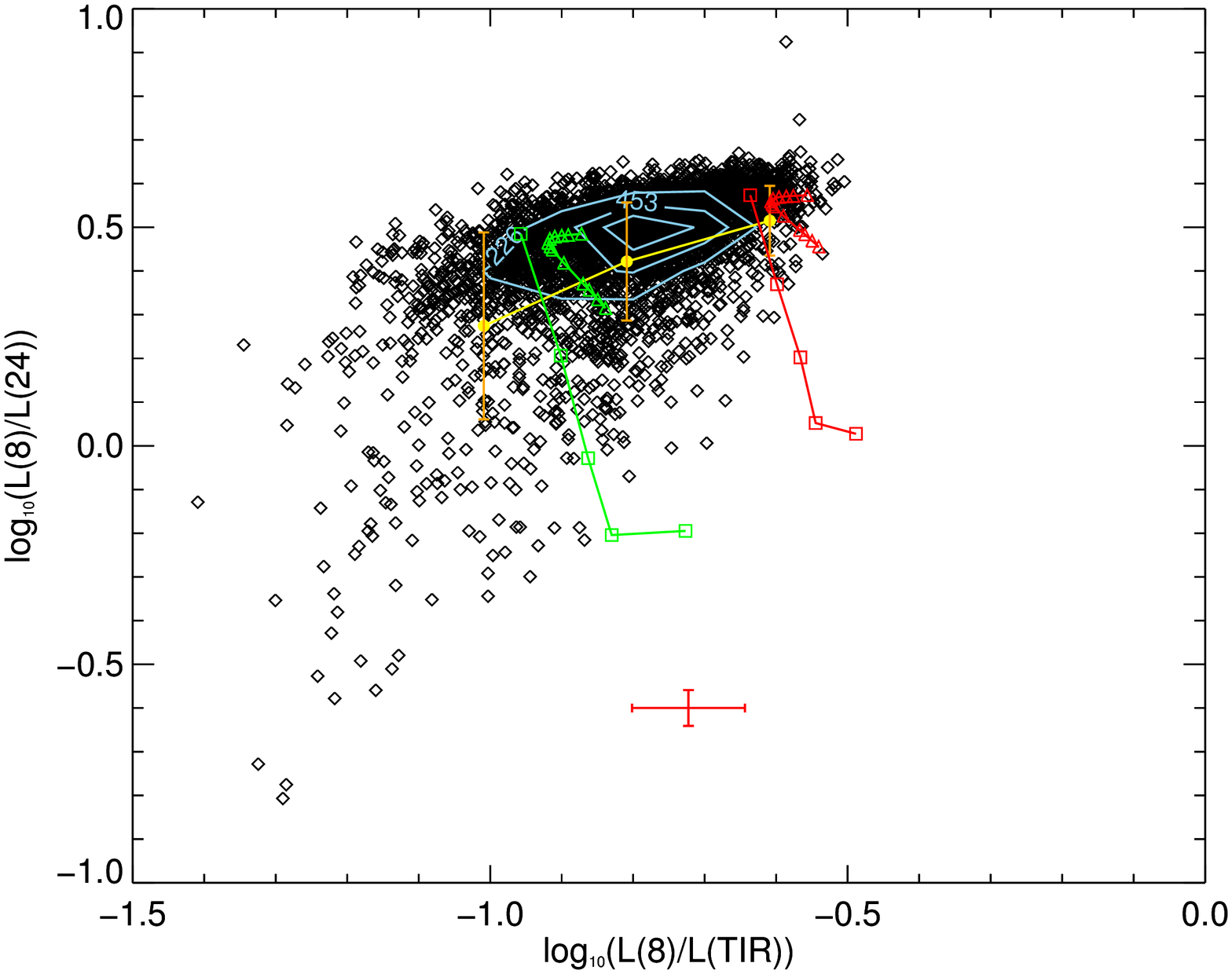}
\caption{The relation between the L(8)/L(24) and the L(24)/L(TIR) and L(8)/L(TIR) ratios. The median values of the data are presented as the yellow points and line, with dispersion bars at each point the median was calculated. The predicted values for the emissivity from \citet{draine07} for the q$_{PAH}$ = 0.046 case (red lines; section 3.2) and the q$_{PAH}$ = 0.025 case (green lines), are plotted for both U$_{min}$ = U$_{max}$ in the range 0.1 - 25 (triangle points) and U$_{min}$ = 0.1 with U$_{max}$ ranging from 0.1 to 10$^{6}$ (square points). The purple box in the left panel demarcates the area used for the calculation of diffuse fractions (see section 4).}
\end{figure}
\indent
Additional insights can be obtained by plotting the L(8)/L(24) ratio as a function of the L(8)/L(TIR) and L(24)/L(TIR) ratios. In this case, the observed data can be compared directly with predictions from models. Figure 6 shows the observed L(8)/L(24) versus L(24)/L(TIR) and L(8)/L(TIR).\\
\indent
The L(8)/L(24) ratio as a function of the L(24)/L(TIR) ratio shows that the 8 $\mu$m luminosity is low relative to the 24 $\mu$m in areas of high 24 $\mu$m luminosity. The second ratio plot as a function of L(8)/L(TIR) confirms this, and implies that 8 $\mu$m emission is under luminous in areas of high star formation. We quantify this statement by calculating the fraction with low L(8)/L(24) within 150 pc of the 119 HII regions from \cite{rel13}. We find that for all pixels below Log(L(8)/L(24)) = 0.15 (about 3.5\% of all pixels above 5-$\sigma$), 60\% of them are located within 150 pc of an HII region. Conversely, when considering pixels with log(L(8)/L(24)) $\ge$ 0.15, only 18\% of them are within 150 pc of one of Rela\~{n}o et al.'s HII regions. These fractions do not change when choosing different  thresholds for Log(L(8)/L(24)), as long as they are $\le$ 0.2. This ratio corresponds to a starlight intensity of U $\approx$ 10$^{2}$ \citep{draine07} for most q$_{PAH}$ values. When analyzing independently the L(8)/L(TIR) and L(24)/L(TIR) as a function of distance from HII regions, we find that while L(8)/L(TIR) is roughly constant or slightly decreasing in value when approaching an HII region, L(24)/L(TIR) shows an increase, thus confirming the L(8)/L(24) trend. This can also be seen in Figure 7, where we plot the fraction of pixels within distance R from the nearest HII region and
with L(8)/L(TIR) below a certain value (X) versus the distance. For L(8)/L(TIR) $<$ 0.063 (log(L(8)/L(TIR) $<$ -1.2, see Figure 6, right
panel), the L(8)/L(TIR) ratio is higher closer to an HII region, implying
that low values of L(8)/L(TIR) tend to cluster close to HII regions. As
X increases, there are two effects: the absolute value of the ratio
increases and tends towards unity (as expected, since now we include
a larger range of values for L(8)/L(TIR)), and the trend as a function
of increasing radius disappears, which is also expected if large
values of L(8)/L(TIR) can be both close and far from HII regions. This is due to the fact that, even in a pixel containing an HII
region, much of the IR emission is from dust located along the
line of sight and not directly associated with the HII region;
only in correspondence of a bright HII region we can expect a
measurable effect on the overall 8 $\mu$m emission from the pixel. This reinforces the conclusions based on L(8)/L(24): low values of 8 $\mu$m surface brightness tend to be closer to HII regions.\\

\begin{figure}
\plotone{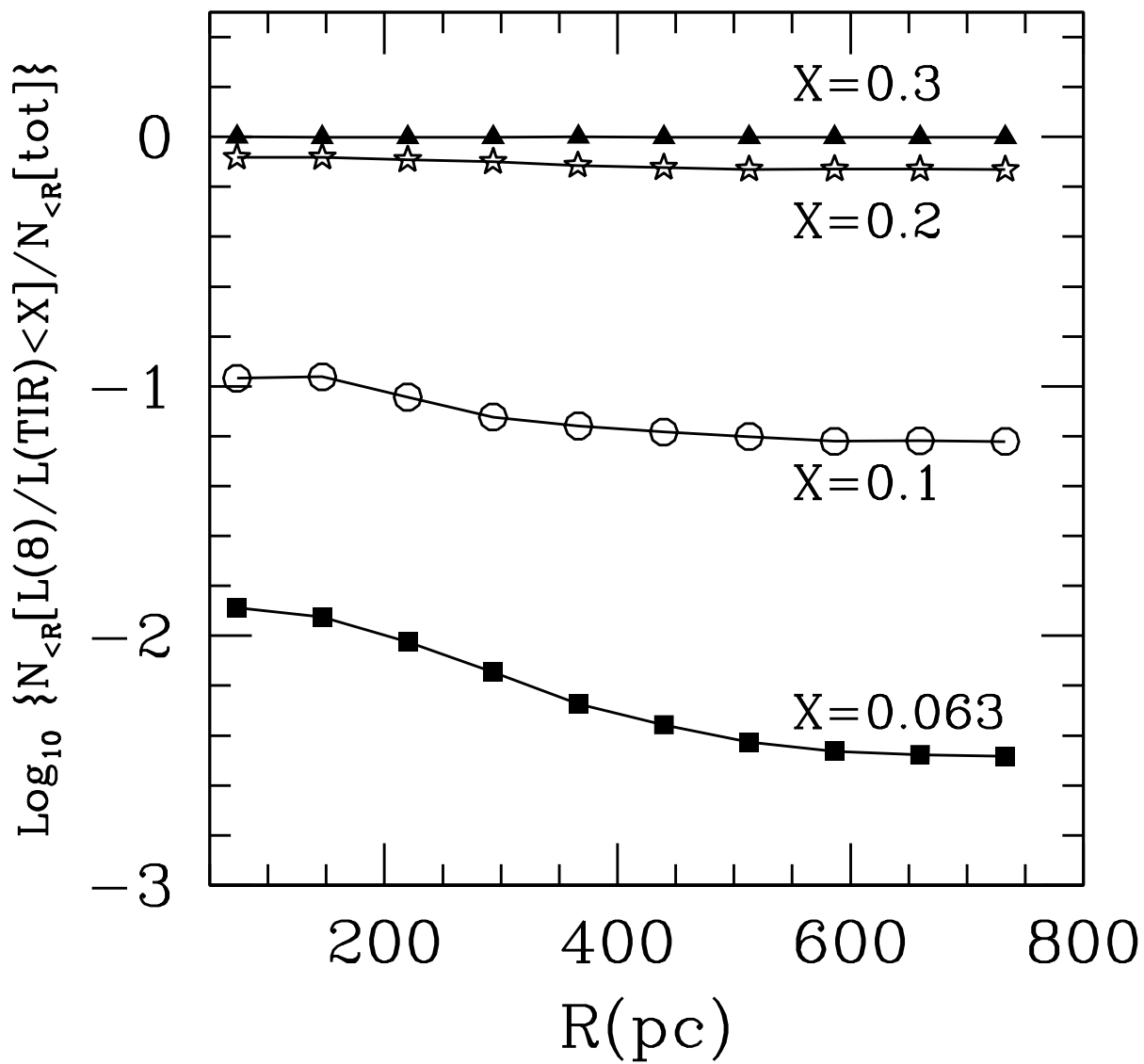}
\caption{The fraction of pixels within distance R with L(8)/L(TIR) below a certain value (X) as a function of distance from the nearest HII region.}
\end{figure}

\indent
The 8 $\mu$m emission is \em{underluminous} \rm relative to expectations from models that include heating from HII regions (exemplified here adopting U$_{max}$ $\approx$ 10$^{6}$ in the \citet{draine07} models, as suggested by those authors). This supports earlier findings and suggestions that the PAH carriers tend to be destroyed in regions of high starlight intensity \citep{helou04, bendo08, pov07, gordon08}.\\

%\begin{figure}
%\plotone{8_24_36.eps}
%\caption{The relation between the L(8)/L(24) and the 3.6 $\mu$m luminosity. The median values of the data are presented as the yellow points and line, with dispersion bars at each point the median was calculated. The predicted values for the emissivity from \citet{draine07} for the q$_{PAH}$ = 0.046 case (red line; section 3.2) are plotted for U$_{min}$ = U$_{max}$ in the range 0.1 - 25.}
%\end{figure}
%\indent
%The L(8)/L(24) ratio shows weak or no correlation as a function of 3.6 $\mu$m luminosity (Figure 8). The 8 $\mu$m emission can still be correlated with old stellar populations, as will be shown in the next section (see Figures 10 and 12). 

\subsection{Relationships Between 8 $\mu$m and 250 $\mu$m Emission}

\begin{figure}[h]
\epsscale{1.1}
\plottwo{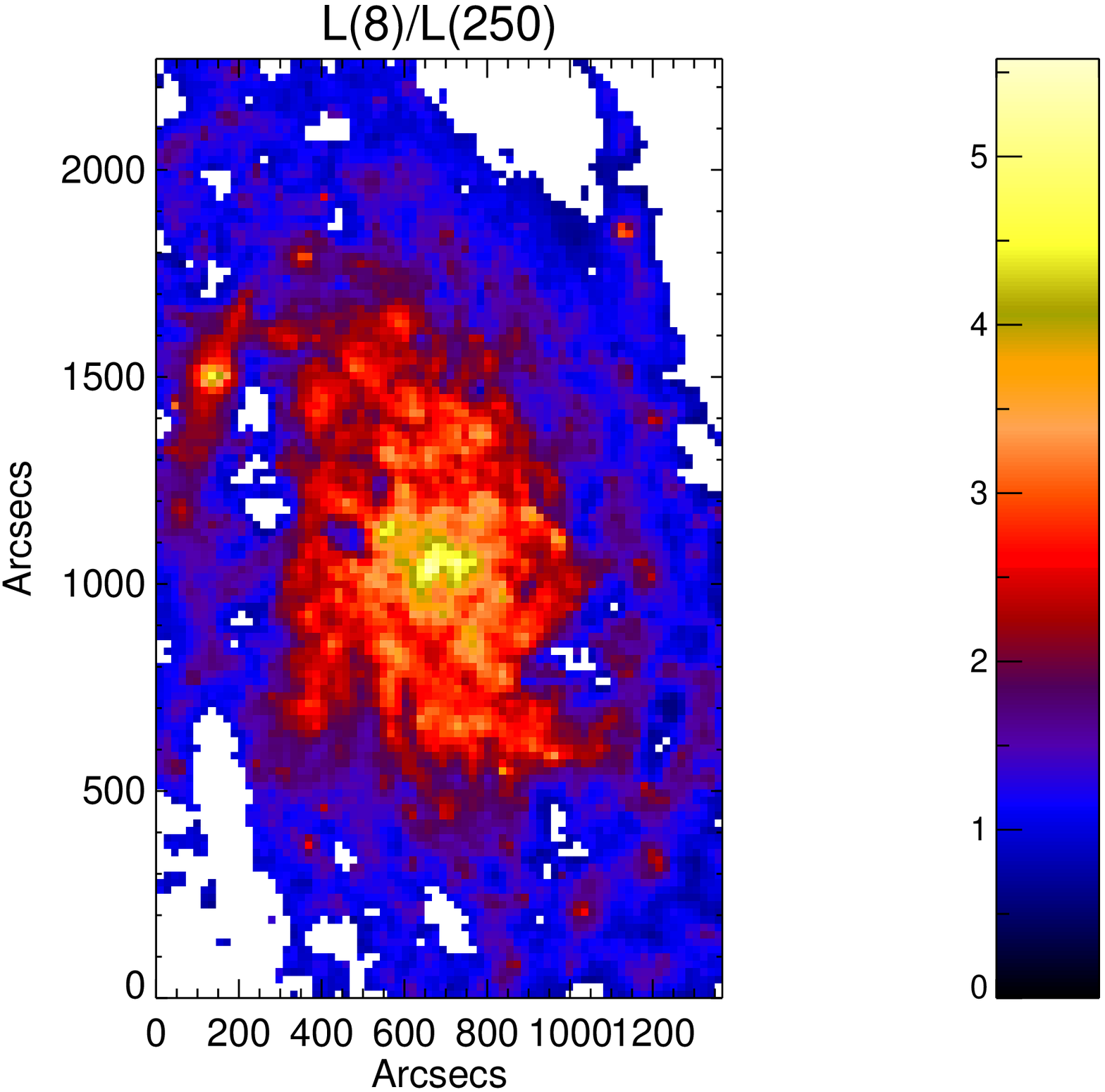}{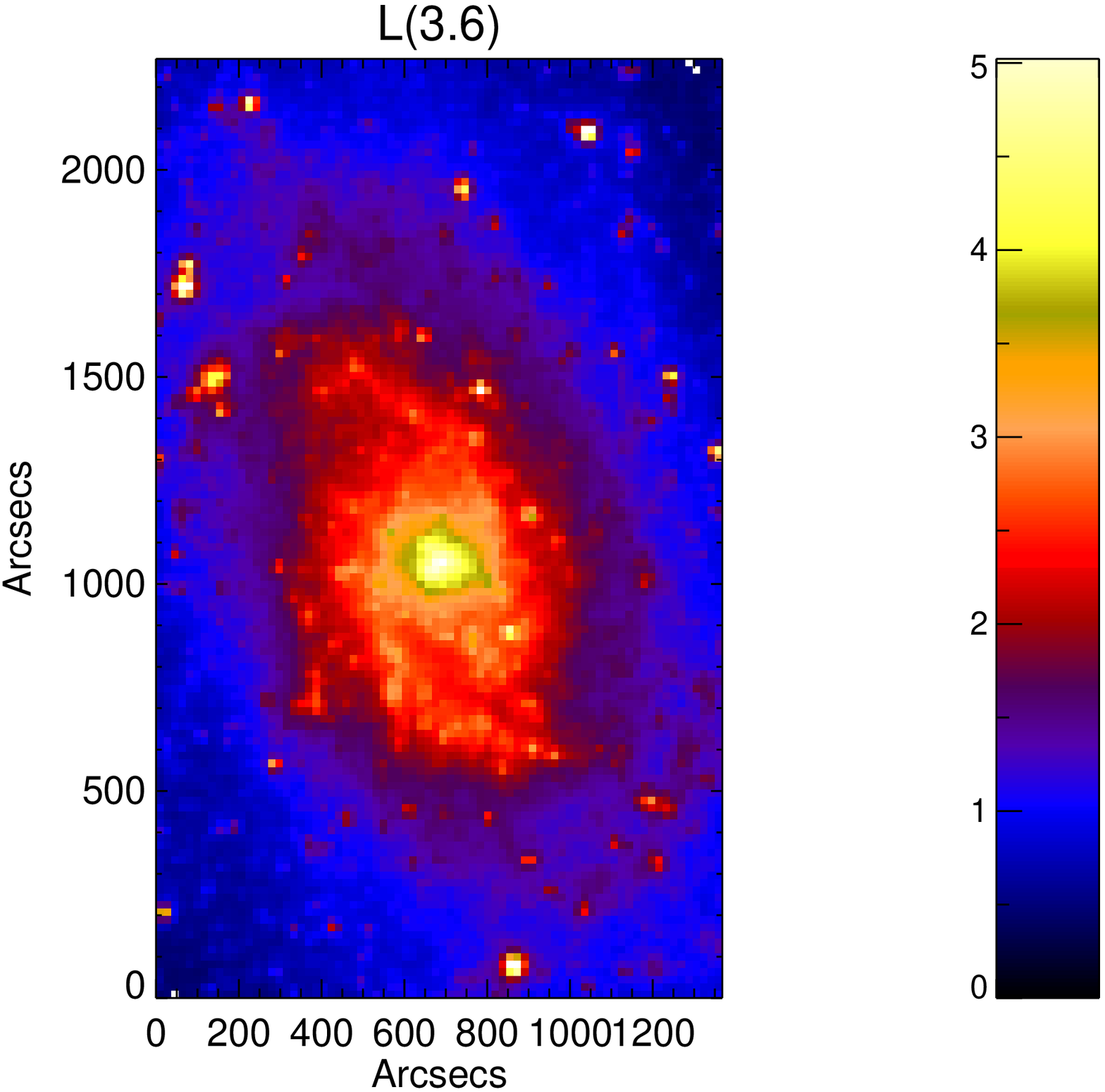}
\caption{The ratio image of L(8)/L(250) luminosity. Resolution is detailed in the caption of Figure 4. For comparison, the 3.6 $\mu$m map is presented on the right; the units of the color bar for the L(3.6) image are 10$^{40}$ erg s$^{-1}$ kpc$^{-2}$. The similar radial trends inherent in both the L(8)/L(250) ratio and 3.6 $\mu$m images suggests that the 8 $\mu$m emission traces the same old stellar population as the 3.6 $\mu$m emission.}
\end{figure}

\begin{figure}
\epsscale{0.8}
\plotone{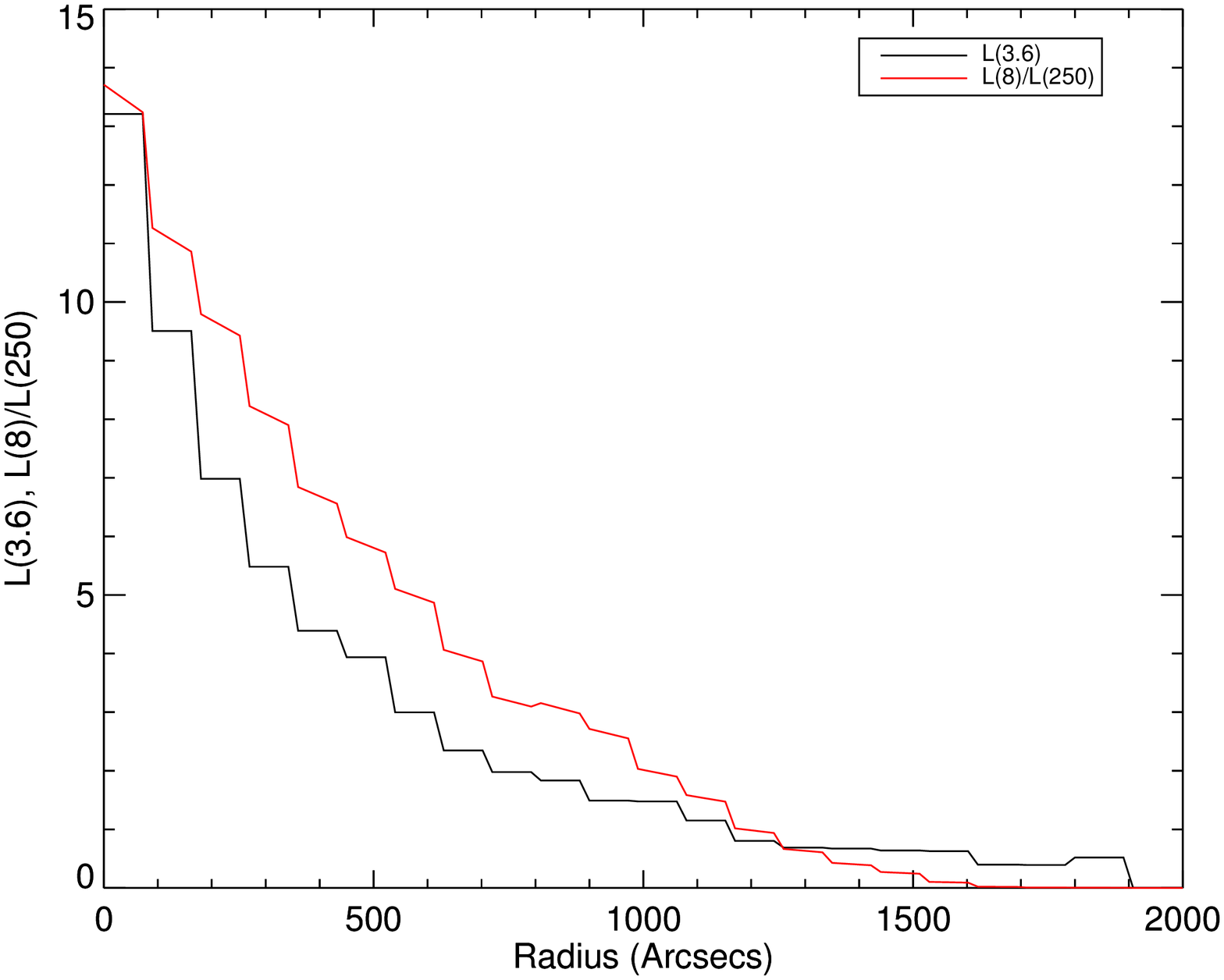}
\caption{The azimuthally averaged galactocentric trends
for both L(8)/L(250) and L(3.6), where L(8)/L(250) have been scaled
in the vertical direction by 3 to match the central value of L(3.6).
This plot aims at highlighting the similarities in the two trends, as
also shown by the two panels of Figure 8. The values for the 3.6 $\mu$m emission are in black, while the ratio is represented by the red line. Units for the 3.6 $\mu$m luminosity are 10$^{40}$ erg s$^{-1}$ kpc$^{-2}$.}
\end{figure}
\noindent
An interesting comparison of the L(8)/L(250) ratio with the old stellar population, traced with the 3.6 $\mu$m emission, suggests that the PAH actually traces old stars. Figure 8 shows the ratio image next to the 3.6 $\mu$m image; which shows strong similarity. Figure 9 compares the galactocentric trends more quantitatively through elliptical summations of signal as a function of radius. The presence of a galactocentric gradient in L(8)/L(250) suggests the
L(8) increases towards the center of M33 faster than L(250) does. Although not
shown, we find a similar result for L(8)/L(TIR), which indicates that the trend in
L(8)/L(250) is not due to changes in the thermal peak of the dust emission. This
trend is likely an indication of a closer correlation of L(8) to the ISRF heating
as traced by the 3.6~$\mu$m emission than to star formation or young stellar
population tracers.

\begin{figure}
\epsscale{1}
\plotone{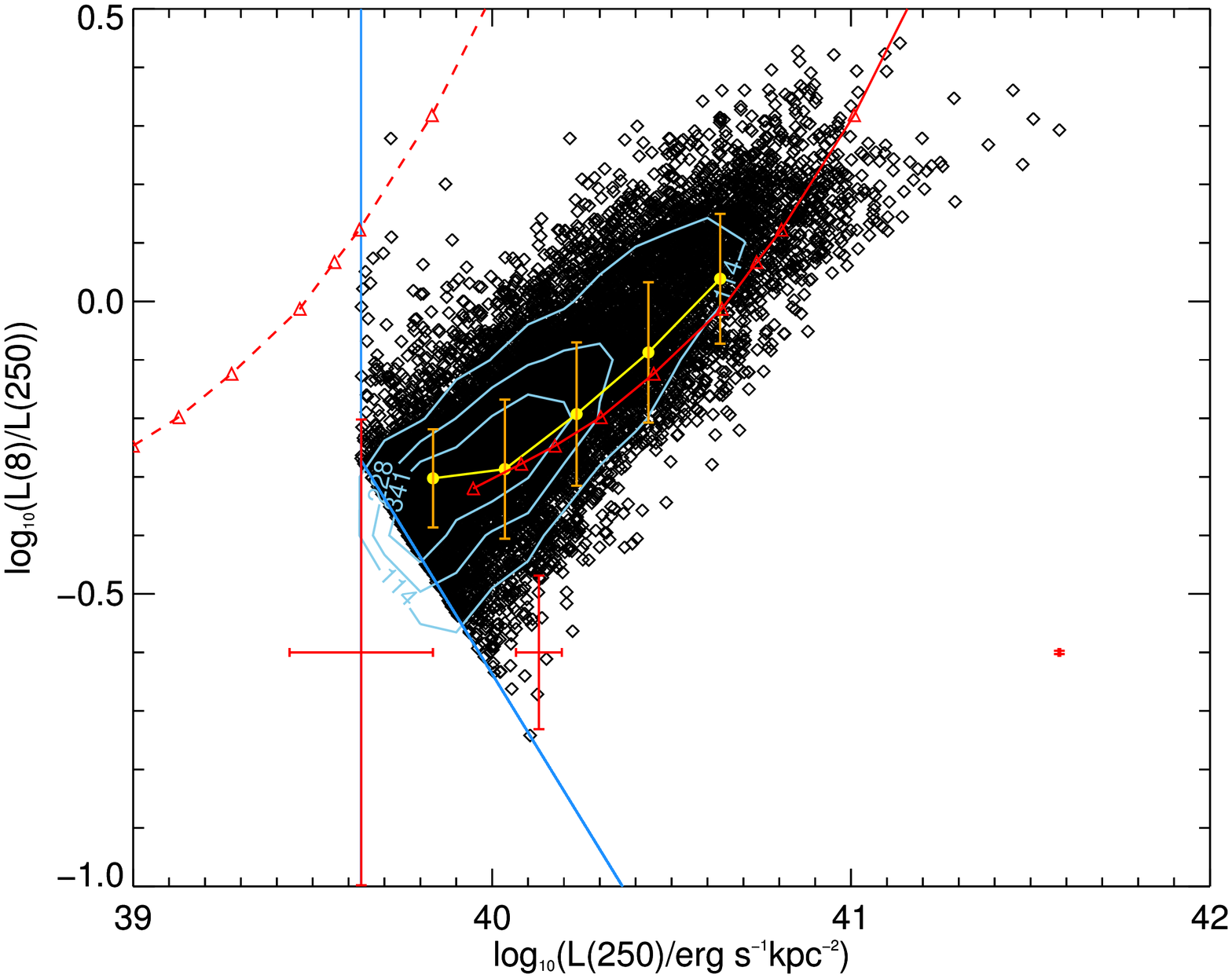}
\caption{The ratio of L(8)/L(250) versus 250 $\mu$m luminosity. Plot symbols and colors are the same as those in Figure 5 for the q$_{PAH}$ = 0.046 model.}
\end{figure}

Figure 10 shows the relation of the L(8)/L(250) ratio with the 250 $\mu$ m
surface brightness. The plot shows a positive slope, suggesting that the 8 $\mu$ m
brightness increases more than the 250~$\mu$m brightness as the latter becomes
more luminous, with no evidence of a change in trend (a decrease), as found in the
L(8)/L(24) versus L(24) scatter plot. \\
\indent
The slope obtained in our data is higher than that obtained by Bendo et al. (2008), but this is explained by the fact that our resolution allows for more accurate tracing of cold dust. We also use a longer wavelength tracer of cold dust then these authors (250 $\mu$m as opposed to 160 $\mu$m) implying that we are further along the Rayleigh-Jeans tail of the dust emission. The range spanned by the data is fully accounted for by the diffuse ISRF with U$_{min}$ in the range 0.1 - 25. From Figure 10, we infer that the correlation of the PAH emission with cold dust is stronger than with star forming regions.\\ 

%\begin{figure}
%\epsscale{1}
%\plotone{8_250_TIR.eps}
%\caption{The ratio of L(8)/L(250) versus TIR luminosity. Plot symbols and colors are the same as those in Figure 5 for the q$_{PAH}$ = 0.046 model.}
%\end{figure}
\indent
This is reinforced by the scatter plot of the L(8)/L(250) ratio as a function of the
TIR luminosity (not shown, but qualitatively similar to Figure 10), which we find
to still show a strong increase of L(8)/L(250) with L(TIR). This reinforces our
earlier conclusion that while the PAH emission increases with the ISRF intensity
more than the 250 $\mu$m emission, this trend is not due to a shift of the thermal
peak of the dust emission to higher temperatures.\\
\indent
A very tight correlation can be found by plotting the L(8)/L(250) ratio as a function of 3.6 $\mu$m luminosity, as in Figure 11. The marked decrease in scatter for L(3.6) $>$ 3.2 $\times$ 10$^{40}$ erg s$^{-1}$ kpc$^{-2}$, by almost a factor of 2, provides further support for the statement that L(8)/L(250) emission is strongly correlated with 3.6 $\mu$m emission, which is a tracer of old stellar populations. A histogram of the residuals of a best fit line through the data quantifies the dispersion (right panel of Figure 11). The blue histogram shows the dispersion of all data points in the left plot, while the red histogram shows the dispersion of all data above L(3.6) = 3.2 $\times$ 10$^{40}$ erg s$^{-1}$ kpc$^{-2}$. Gaussian fits to the histograms give FWHMs that decrease from 0.19 for all data to 0.11 for the bright L(3.6) points, almost a factor of two decrease. \\

\begin{figure}[h]
\epsscale{1}
\plottwo{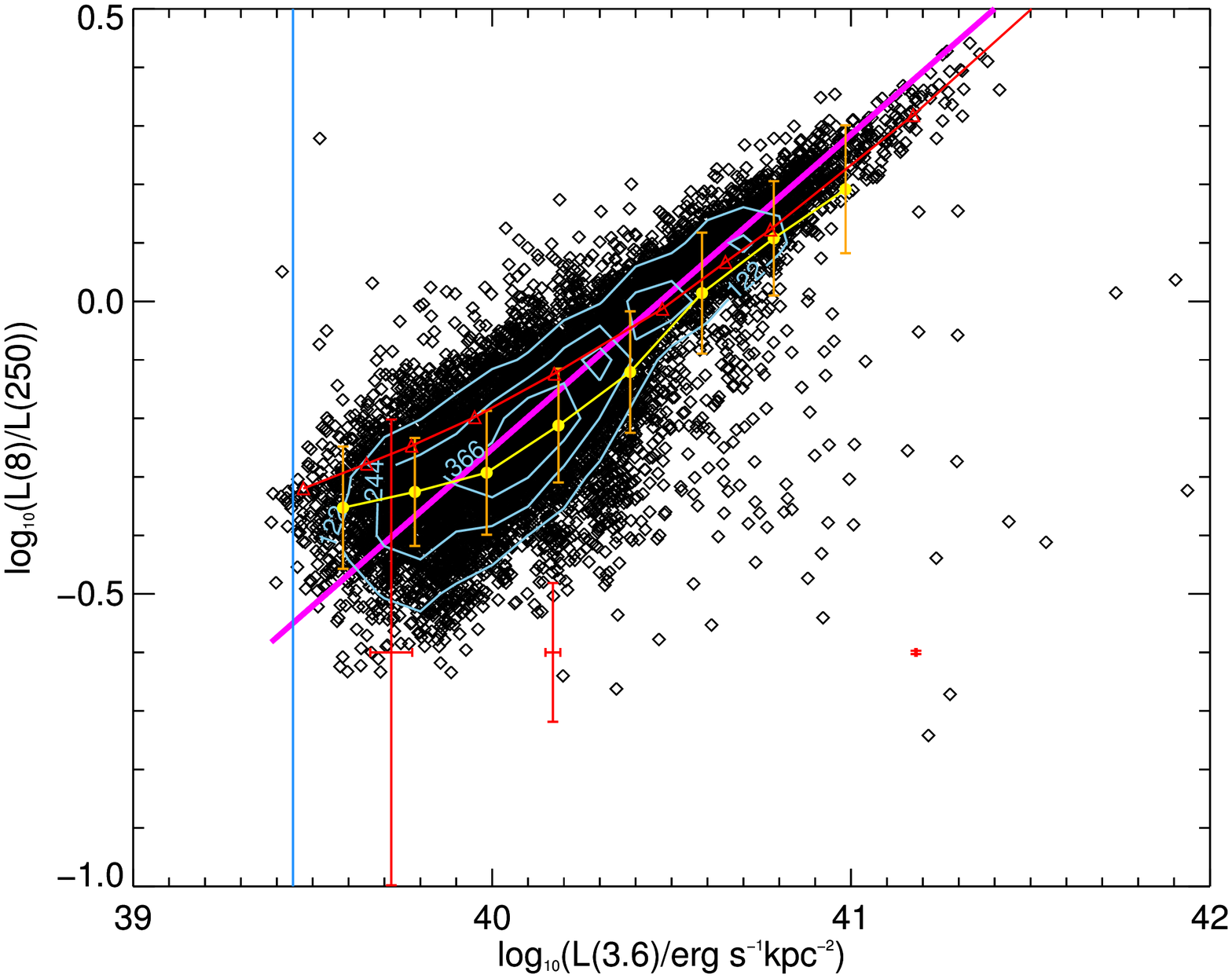}{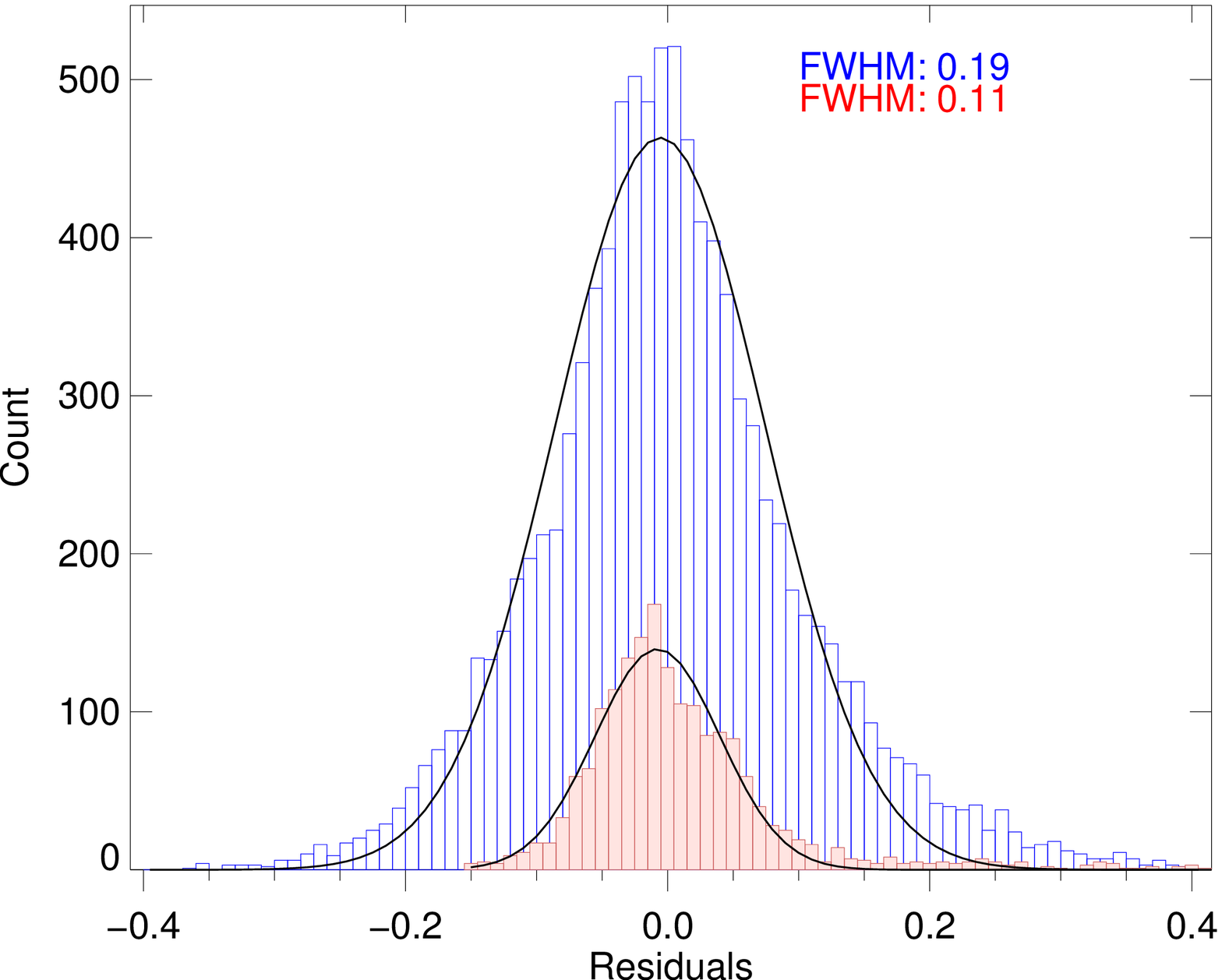}
\caption{The ratio of L(8)/L(250) versus 3.6 $\mu$m luminosity. Plot symbols and colors are the same as those in Figure 5 for the q$_{PAH}$ = 0.046 model. The histogram on the right shows the spread of the data from the best fit line in magenta. The blue histogram is for all the data points to the left, while the red histogram is for data points above L(3.6) = 3.2 $\times$ 10$^{40}$ erg s$^{-1}$ kpc$^{-2}$.}
\end{figure}
\indent
When we plot the L(8)/L(250) ratio as a function of the combined luminosity of H$\alpha$ and 24 $\mu$m emission, which is an accurate SFR indicator \citep{calz07}, as in Figure 12, we still find a correlation, but not as good as that found in Figure 11, with the 3.6 $\mu$m emission. This is evidenced by comparing the histograms of the residuals.  When comparing the histograms in Figure 11 and 12, we notice a markedly lower dispersion in L(8)/L(250) when plotted as a function of L(3.6) than as a function of L(H$\alpha$) + 0.03L(24). This is particularly true in the bright regime, where the dispersion relative to L(3.6) is half of that relative to the SFR tracer. In order to verify that the tighter relation with the 3.6 $\mu$m image is not a by-product of the presence of the 3.3 $\mu$m PAH emission in the IRAC1 band, we perform the same analysis using the PAH - free IRAC2 band (4.5 $\mu$m), recovering the same result. We consider this further support of the better association of the 8 $\mu$m emission with the heating by evolved stars.\\

\begin{figure}[h]
\epsscale{1}
\plottwo{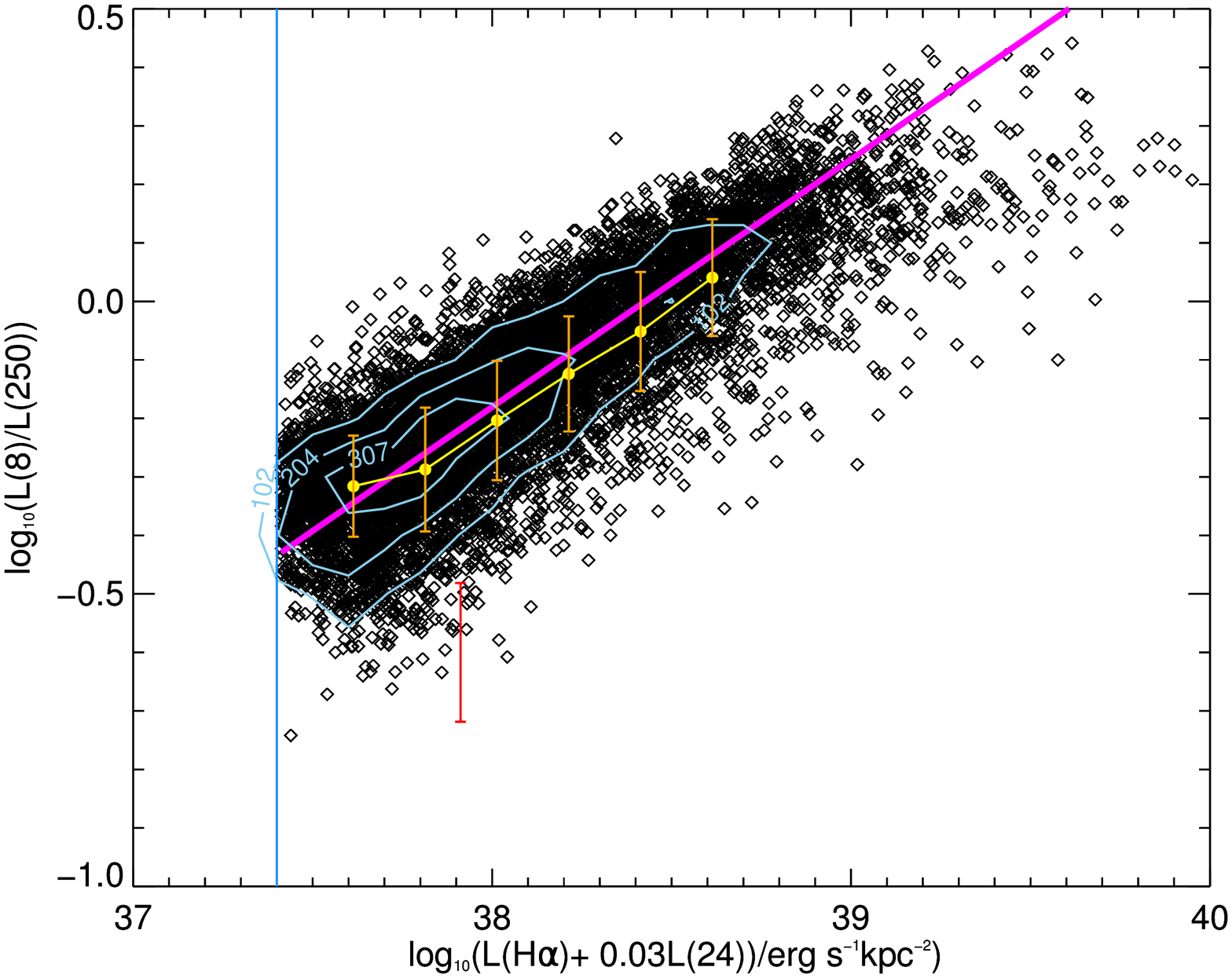}{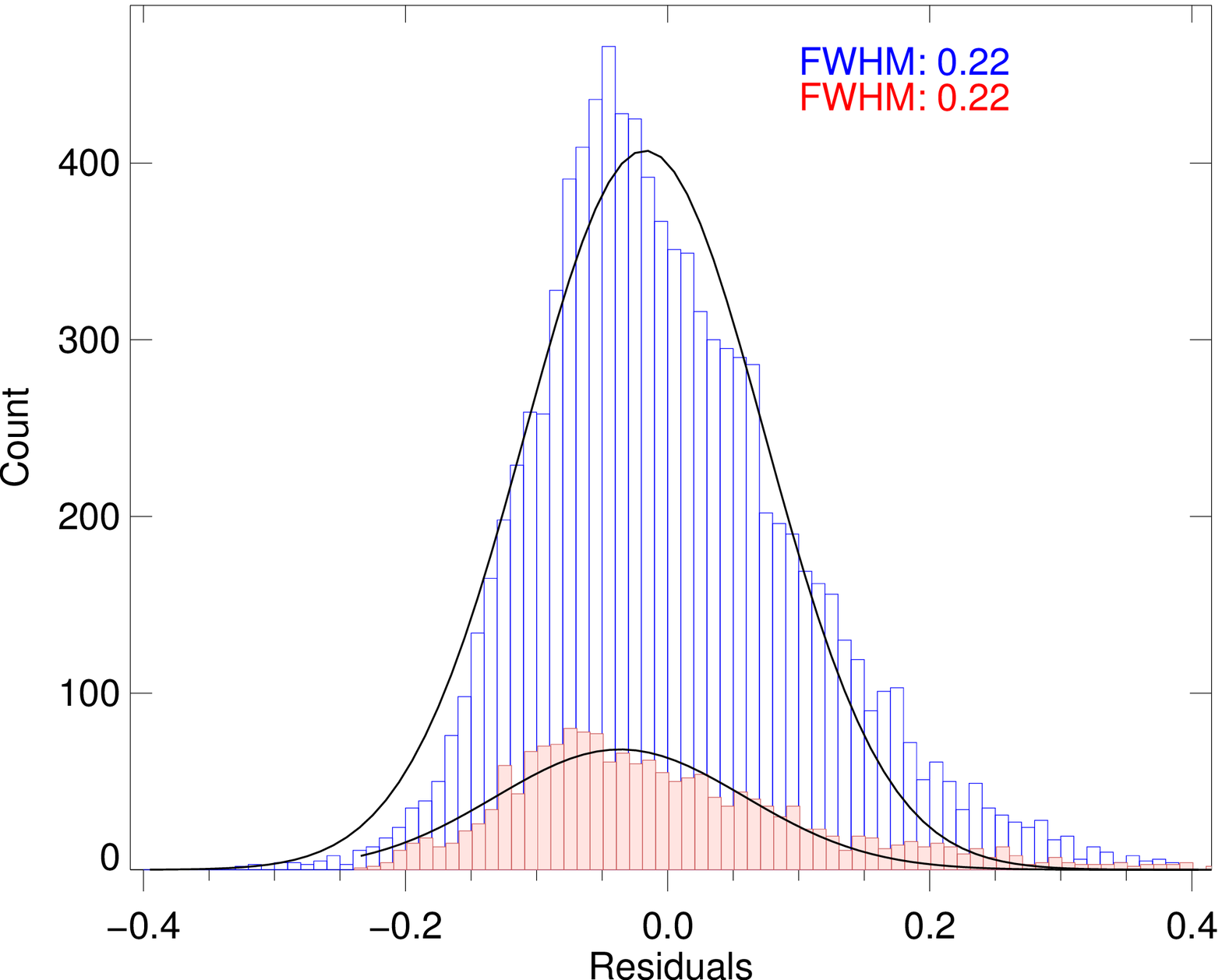}
\caption{The ratio of L(8)/L(250) versus a combination H$\alpha$ and 24 $\mu$m luminosity. Plot symbols and colors are the same as those in Figure 5 for the q$_{PAH}$ = 0.046 model. The histogram on the right shows the spread of the data from the best fit line in magenta. The blue histogram is for all the data points to the left, while the red histogram is for data points above L(H$\alpha$) + 0.03L(24) = 3.2 $\times$ 10$^{38}$ erg s$^{-1}$ kpc$^{-2}$.}
\end{figure}

\section{Diffuse Fractions}
\noindent
We use the plot of L(8)/L(24) versus L(24)/L(TIR) (Figure 6, left) to attempt to separate bins containing dust mainly heated by the diffuse ISRF from those mainly heated by HII regions. We define the diffuse regions as marked by the boundaries -1.65 $\le$ log$_{10}$($\frac{L(24)}{L(TIR)}$) $\le$ -1.1 (or -1.2) and 0.3 $\le$ log$_{10}$($\frac{L(8)}{L(24)}$) $\le$ 0.7. While these choices are somewhat arbitrary, we select the region to be symmetric relative to the mean value of L(8)/L(24) (in log scale) at low L(24)/L(TIR), and to be contained within the L(24)/L(TIR) ratio values that mark a relatively constant (and high) L(8)/L(24) ratio. With this choice, we find the fractions of L(8) and L(24) associated with diffuse emission are 80 - 59\% and 60 - 43\%, respectively. The range in the fractions is a result of applying the different limits (-1.1 or -1.2) on the selection of data. While these values are uncertain, we always find that the fraction of diffuse L(8) is 33 - 42\% higher than the fraction of diffuse L(24) independently of how we select the region of diffuse emission. Xilouris et al. (2012) use a different approach from the one presented here, based
purely on morphological arguments(decomposing the total emission in the diffuse cold
dust in the disk and the dust associated with star forming regions in the spiral arms); yet, they find
a similar trend to ours for the fraction of the diffuse emission at 8~$\mu$m to be 29\%,
higher than the value of 21\% at 24~$\mu$m that those authors derive.
Our and Xilouris et al.Õs results are slightly different than those of Verley
et al. (2009), who derive comparable fractions, 60\% to 80\% of diffuse emission
at both 8~$\mu$m and 24~$\mu$m. However, our and these authorsÕ results are not
necessarily inconsistent when taking into account the different approach used
in that work. If the average among galaxies of the diffuse L(24) is in the range 20 - 40\% \citep{calz07, ken09, leroy12}, the average diffuse L(8) is in the range 30 - 60\%, in rough agreement with the results of \citet{crock13} for NGC 628.\\

\section{Conclusion}
\noindent
In this paper, we have investigated the relationships between the 8 $\mu$m emission and emission in other wavelengths that are correlated to old stellar populations, star formation, and cold dust. We have also compared our data with the predictions of the models from \citet{draine07}. Ratios with 24 $\mu$m and TIR luminosity show the 8 $\mu$m emission to become underluminous in areas of strong 24 $\mu$m emission. Taking the 24 $\mu$m emission to be the product of heating of dust by young stellar radiation, this implies that 8 $\mu$m emission is not a close tracer of young stellar populations, and the carriers responsible for its emission are possibly being destroyed by the intense radiation fields of star forming regions. Furthermore, the behavior of most ($\ge$ 80\%) 8 $\mu$m emission can be explained using only the U$_{min}$ = U$_{max}$ models, which is valid for diffuse ISRF. We also find that the values of U$_{min}$ that account for the
observed 8~$\mu$m emission are in the range 0.1 - 25, consistent with
the range found by Draine et al. (2014) for the galaxy M31. When applying a model that includes heating from HII regions, the observed 8 $\mu$m values are lower than expected, supporting other authors' conclusions that the carriers responsible for 8 $\mu$m emission are destroyed in intense stellar radiation fields. An alternative explanation to the destruction of the 8 $\mu$m carriers
is to make the PAHs more neutral, suppressing the 8 $\mu$m emission and
shifting power to the 11-12 $\mu$m features\citep{sand12}. Mid-IR spectroscopy may
discriminate among these two scenarios, although we tentatively give preference to the
PAH destruction interpretation.\\
\indent
Ratios of 8 $\mu$m with 250 $\mu$m emission show a strong correlation with both the TIR and 3.6 $\mu$m luminosities. Both 3.6 $\mu$m emission and L(8)/L(250) show a similar galactocentric radial trend, suggesting the 8 $\mu$m emission may be more connected to old stellar populations than star formation. Further, the L(8)/L(250) as a function of 3.6 $\mu$m luminosity shows the tightest relation among those presented in this paper, especially at high luminosity.\\
\indent
We also derive the fraction of 8 $\mu$m and 24 $\mu$m emission heated either by the ISRF or HII regions. The fraction of L(24) associated with diffuse emission is 60 - 43\%, while the fraction of L(8) is nearly 33 - 42\% higher, with 80 - 59\% of the emission coming from heating by the ISRF.\\
\indent
The results of this study, which support findings by other authors, but at the exquisite spatial resolution enabled by the proximity of M33, suggest that the 8 $\mu$m luminosity should not be used as a proxy for measuring and locating star formation in galaxies. Emission in this wavelength is shown in M33 to be more correlated with cold dust or old stellar populations than with star formation.

\section{Acknowledgements}
\noindent
This work is based on observations made with Herschel, a European Space Agency Cornerstone Mission with significant participation by NASA, using the
PACS and SPIRE instruments. Partial support for this work was provided by NASA through an award issued by JPL/Caltech.\\
\indent
The Herschel spacecraft was designed, built, tested, and launched under a contract to ESA managed by the Herschel/Planck Project team by an industrial consortium under the overall responsibility of the prime contractor Thales Alenia Space (Cannes), and including Astrium (Friedrichshafen) responsible for the payload module and for system testing at spacecraft level, Thales Alenia Space (Turin) responsible for the service module, and Astrium (Toulouse) responsible for the telescope, with in excess of a hundred subcontractors.\\
\indent
PACS has been developed by a consortium of institutes led by MPE (Germany) and including UVIE (Austria); KU Leuven, CSL, IMEC (Belgium); CEA, LAM (France); MPIA (Germany); INAF-IFSI/OAA/OAP/OAT, LENS, SISSA (Italy); IAC (Spain). This development has been supported by the funding agencies BMVIT (Austria), ESA-PRODEX (Belgium), CEA/CNES (France), DLR (Germany), ASI/INAF (Italy), and CICYT/MCYT (Spain).\\
\indent
SPIRE has been developed by a consortium of institutes led by Cardiff University (UK) and including Univ. Lethbridge (Canada); NAOC (China); CEA, LAM (France); IFSI, Univ. Padua (Italy); IAC (Spain); Stockholm Observatory (Sweden); Imperial College London, RAL, UCL-MSSL, UKATC, Univ. Sussex (UK); and Caltech, JPL, NHSC, Univ. Colorado (USA). This development has been supported by national funding agencies: CSA (Canada); NAOC (China); CEA, CNES, CNRS (France); ASI (Italy); MCINN (Spain); SNSB (Sweden); STFC, UKSA (UK); and NASA (USA).\\
\indent
This research has made use of the NASA/IPAC Extragalactic
Database (NED) which is operated by the Jet Propulsion
Laboratory, California Institute of Technology, under contract
with the National Aeronautics and Space Administration.\\

\appendix 
\section{The Relation Between Observed Luminosity Density $\Sigma_{\nu}$ and Internal Energy Density $u_\nu$}
\noindent
We want to relate the energy density measured within a galaxy to the luminosity 
surface density measured when viewing a galaxy from the outside, as projected on the sky. This exercise 
enables us to relate the energy density of ISRF of the Milky Way, which we measure from within the galaxy, 
to the luminosity density of the ISRF of external galaxies.\\
\indent
For this, we adopt a number of simplifying assumptions, that enable us to perform an analytic calculation:
\begin{itemize}
\item the galaxy is a disk, approximated plane-parallel; 
\item the density of stars is everywhere proportional to the density of dust, i.e., the two are homogeneously mixed; 
\item the dust is purely absorbing (this assumption is justified if the scattering out of the line of sight roughly compensates 
the scattering into the line of sight). 
\end{itemize}
Let $\tau_0$ be the dust optical depth normal to the disk.
Let $n_\star$ be the density of stars, $L_{\star,\nu}$ the luminosity per
unit frequency of one star, $n_d$ be the density of dust, and $\sigma_d$ be the absorption
cross section per dust grain.
The equation of radiative transfer is
\begin{equation}
\frac{dI_\nu}{ds} = - n_d\sigma_d I_\nu + \frac{n_\star L_{\star,\nu}}{4\pi}
\end{equation}
\begin{equation}
\frac{dI_\nu}{d\tau} = -I_\nu + \frac{n_\star L_{\star,\nu}}{4\pi n_d\sigma_d}
~~,
\end{equation}
where $d\tau = n_d\sigma_d ds$.\\
\indent
Because the source function $n_\star L_{\star,\nu}/4\pi n_d\sigma_d$ is
constant, this is easily integrated:
\begin{equation}
I_\nu = S_\nu\left[1-e^{-\tau}\right]
~~,
\end{equation}
where $\tau$ is the optical depth from the point of interest to $\infty$, and
the source function is
\begin{equation}
S_\nu\equiv\frac{n_\star L_{\star,\nu}}{4\pi n_d\sigma} ~~.
\end{equation}

The above equations will now be used to directly relate the energy density measured at some given point within the 
disk to the disk surface brightness. From any point within the disk,  the optical depth normal
to the disk is $z\tau_0$ in one direction, and  $(1-z)\tau_0$ in the opposite direction,
where $0<z<1$.
The specific energy density at this point is
\begin{equation}
u_\nu = \frac{1}{c} \int_{0}^{\pi} 2\pi\sin\theta d\theta ~I_\nu(\theta)
~~,
\end{equation}
where $\theta$ is measured relative to the disk normal.
\begin{eqnarray}
u_\nu &=& \frac{4\pi S_\nu}{c} F(\tau_0,z)\\
F(\tau_0,z) &\equiv&
\int_0^1 d\mu \left[ 1- 
\frac{1}{2}\left(e^{-z\tau_0/\mu}+e^{-(1-z)\tau_0/\mu}\right)\right]
~~.
\end{eqnarray}
Viewed from outside the disk, with inclination $i$, the intensity is
\begin{equation}
[I_\nu]_i = S_\nu (1-e^{-\tau_0/\cos i}) ~~,
\end{equation}
and the apparent specific luminosity surface density (projected on the
plane of the sky) is
\begin{equation}
\Sigma_{L,\nu} = 4\pi[I_\nu]_i = 4\pi S_\nu (1-e^{-\tau_0/\cos i}) ~~.
\end{equation}
Thus we can relate the energy density within the disk to the
disk surface brightness
\begin{equation}
u_\nu = \frac{4\pi[I_\nu]_i}{c}
\frac{F(\tau_0,z)}{1-e^{-\tau_0/\cos i}}
= \frac{\Sigma_{L,\nu}}{c}\frac{F(\tau_0,z)}{1-e^{-\tau_0/\cos i}} ~~.
\end{equation}

We now specialize our derivation to the specific case of M33. 
Figure~13 shows the ratio $F/(1-e^{-\tau_0/\cos i})$
as a function of $\tau_0$, for $z=0.5$ (midplane), $z=0.1$ (or $0.9$),
and $z=0.01$ (or $0.99$),
for the estimated inclination $i=56^\circ$ of M33.

%%%%%%%%%%%%%%%%%%%%%%%%%%%%% f1 %%%%%%%%%%%%%%%%%%%%%%%%%%%%%%%%%%%%%%
\begin{figure}[t]
\begin{center}
\plottwo{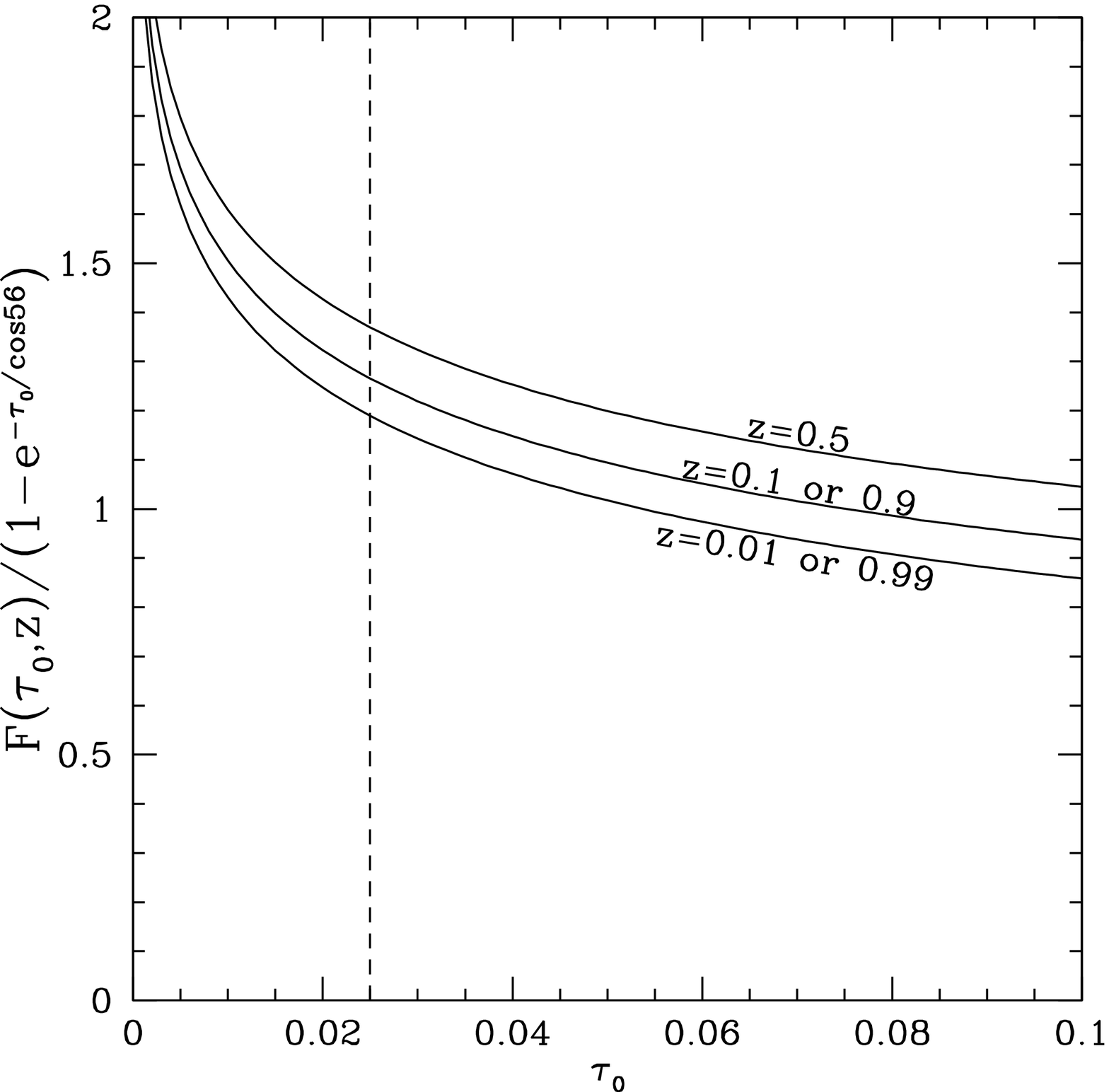}{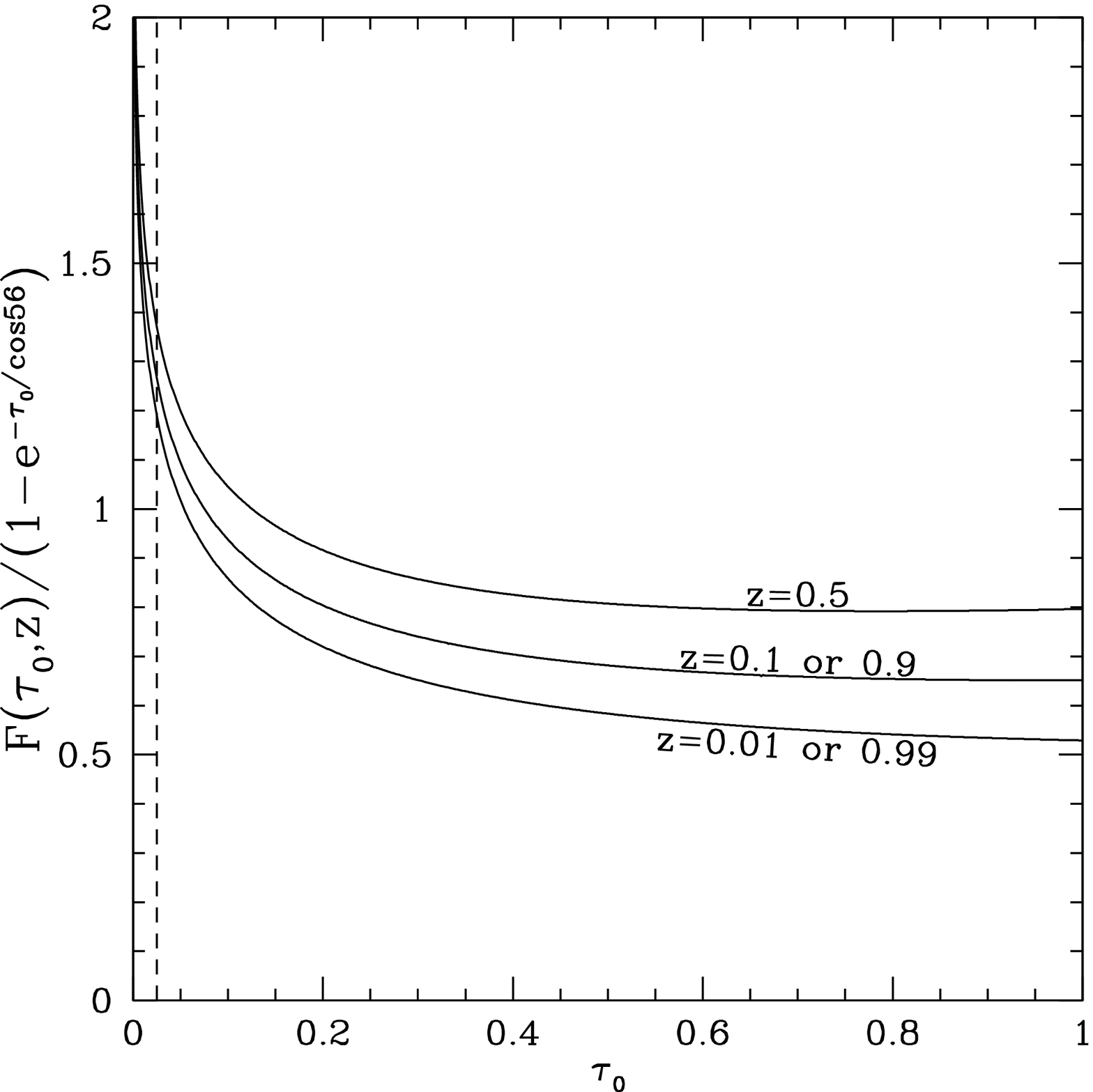}
\caption{
         The ratio $F(\tau_0,z)/(1-e^{-\tau_0/\cos i})$ for the inclination
         $i\approx 56^\circ$ of M33, as a function of
         $\tau_0$, the full-thickness optical depth normal to the disk.
         We see that for $\tau_0\approx 0.025$ estimated for M33 at
         $3.6\micron$, the ratio $F/(1-e^{-\tau_0/\cos i})\approx 1.35$
         within the disk.  The ratio varies only between $1.4$ and $1.2$
         as we move from the midplane to the surface of the disk.
        }
\end{center}
\end{figure}
%%%%%%%%%%%%%%%%%%%%%%%%%%%%% end f1 %%%%%%%%%%%%%%%%%%%%%%%%%%%%%%%%%%%%%

At 3.6$\micron$, MW dust has $n_d\sigma_d/n_H=2.26\times10^{-23}~cm^2/H$.
Applying this to the M33 case, with  metallicity $\sim$0.5$\times$solar and taking as 
an example $N_H\approx4\times10^{21}~cm^{-2}$, we estimate 
$\tau_0(3.6\micron)=0.5\times2.26\times10^{-23}\times4\times10^{21}\times\cos(56^\circ)
\approx 0.025$. From Figure~13 we see that $F/(1-e^{-\tau_0/\cos i})\approx 1.3$ for this 
value of $\tau_0$. Thus at 3.6$\micron$ we estimate 
$\nu u_\nu \approx 1.3\times(1/c)\nu \Sigma_{L,\nu}$.
For the range of H column densities observed in M33, $N_H\sim4\times10^{20}-6\times10^{21}~cm^{-2}$,
the range of $F/(1-e^{-\tau_0/\cos i})$ goes from roughly 1.2 to 1.9, with a corresponding change 
in the relation between $\nu u_\nu$ and $\nu \Sigma_{L,\nu}$. Taking a mean value 
$F/(1-e^{-\tau_0/\cos i})\sim$1.5 and assuming that the starlight spectrum in M33 is the same as 
the local ISRF,  we get:
\begin{equation}
(\nu\Sigma_{L,\nu})_{3.6\micron} = \langle U \rangle\times 1.922\times 10^{40}~erg~s^{-1}~kpc^{-2}, 
\end{equation}
for the local MW ISRF value  
$(\nu u_\nu)_{3.6\micron}\approx 1\times10^{-13}~erg~cm^{-3}$ (see, e.g.,
Fig.\ 12.1 of Draine 2011), and using $\langle U \rangle$ for scaling the radiation 
intensity. We finally include an additional scaling factor $f$ to account for the variations in the 
gas column density ($f\sim[0.75;1.25]$):
\begin{equation}
L(3.6)= (\nu\Sigma_{L,\nu})_{3.6\micron} = \langle U \rangle\times f\times 1.922\times 10^{40}~erg~s^{-1}~kpc^{-2},
\end{equation}
as used in this work.

The above treatment assumes the dust and stars to both be smooth plane-parallel
distributions, with dust density strictly following stellar density.
In a real star-forming galaxy, star-forming regions will have enhanced
values of stellar density/dust density, and the stars will also be
clumped.  Hence,  Equation~A12 should be regarded as only a rough estimate for the
starlight intensity heating the dust in that pixel.

\end{document}